\documentclass[aps,prd,reprint,superscriptaddress,longbibliography,nofootinbib,floatfix]{revtex4-2}
\usepackage{CJKutf8}
\usepackage{booktabs}
\usepackage{graphicx}
\usepackage{amsmath,amssymb,amsthm}
\usepackage{mathrsfs}
\usepackage{orcidlink}
\hypersetup{hidelinks}
\numberwithin{equation}{section}

\theoremstyle{definition}

\theoremstyle{remark}

\begin{document}

\title{Fermion mass hierarchies in a three-generation Clifford model}

\author{Ye Zhou (\begin{CJK*}{UTF8}{gbsn}周烨\end{CJK*})\orcidlink{0009-0004-7050-7736}}
\email{ye.zhou.horizon@gmail.com}
\affiliation{Independent Researcher, Kunshan, Jiangsu, China}
\author{Niels Gresnigt\orcidlink{0000-0002-1993-3025}}
\email{niels.gresnigt@xjtlu.edu.cn}
\affiliation{Department of Physics, School of Mathematics and Physics, Xi’an Jiaotong-Liverpool University, 111 Ren’ai Road, Suzhou 215123, China}

\begin{abstract}
We examine whether an algebraic model of three fermion generations based on the complex Clifford algebra $\mathbb C\ell(10)$, in which an $S_3$ family symmetry relates the generations, can reproduce the observed mass hierarchies of the charged leptons and down quarks. Taking the internal Gram matrix as the fermion kinetic metric, we show that the original Yukawa interaction, with complex coefficients, gives precisely those ordered mass triples $0\leq m_1\leq m_2\leq m_3$ for which $m_3\leq m_1+2m_2$. This restriction excludes the observed hierarchies in both sectors and persists when the right-acting Higgs operator is allowed to range over the full Clifford algebra.

This obstruction motivates a second Yukawa contraction in which a projector, constructed from the difference between an internal charge operator and its family average, acts from the left. The resulting interaction preserves the gauge and family symmetries and uses the same three Higgs doublets. With both contractions present, there is a fixed coupling ratio at which the two successive mass ratios in a single sector become arbitrarily large along a real vacuum path. Separately, a rational interval calculation establishes that one real Higgs vacuum, with different coupling ratios in the two sectors, reproduces the chosen running masses of the charged leptons and down quarks. A softly broken three-doublet potential realises this vacuum as a global minimum at tree level. The additional neutral scalars have flavour-changing couplings, and their exchange at a common mass of $2000\,\mathrm{TeV}$ satisfies the neutral-meson mixing criterion and the $\mu\to3e$ contact-interaction test considered here.
\end{abstract}

\maketitle

\section{Introduction}\label{Sec:Intro}

The tau is about seventeen times as heavy as the muon, although the two leptons carry the same gauge charges~\cite{PDG2026}. Can an algebraic model in which an $S_3$ family symmetry relates the three fermion generations account for this hierarchy and for the different hierarchy observed in the down-quark sector? Division and matrix algebras have provided several ways to represent the fermion families~\cite{Dixon2004FamilyReplication,ManogueDray1999,Furey2018Generations,GordingSchmidtMay2020,GillardGresnigt2019}. In the model considered here, the $S_3$ action is inherited from automorphisms of the sedenions~\cite{Brown1967,GresnigtGourlayVarma2023,GourlayGresnigt2024} and leaves the Standard Model gauge generators invariant~\cite{Gresnigt2026Generations,Gresnigt2026Technical}. The development of the model, from the realisation of the sedenion $S_3$ automorphisms in $\mathbb C\ell(8)$ to the extension to $\mathbb C\ell(10)$, is reviewed in Ref.~\cite{Gresnigt2026Review}. The Higgs extension of this construction, formulated in the complex Clifford algebra $\mathbb C\ell(10)$, provides explicit maps from weak-doublet to weak-singlet fermion states~\cite{Gresnigt2026Higgs}.

The Yukawa coefficients are obtained by pairing the Higgs-transformed left-handed states with the right-handed states. The representatives of different generations are nonorthogonal within each chiral sector, so obtaining the physical masses requires a prescription for the fermion kinetic terms~\cite{KingPeddie2004,EspinosaIbarra2004,ChenEtAl2012}. We take the internal Gram matrix as the kinetic metric and use it to canonically normalise the fermion fields. In the original interaction, a linear combination of three neutral Higgs operators acts on the fermion state by right multiplication. Allowing an arbitrary Clifford element to act from the right enlarges the space of mass matrices while leaving the set of attainable spectra unchanged. In both cases, the heaviest mass can be at most three times the next heaviest, a bound exceeded in both the charged-lepton and down-quark sectors.

The bound follows from the form of the original Yukawa interaction and the chosen kinetic metric. Pakvasa and Sugawara used $S_3$ symmetry in both the Higgs--fermion interactions and the scalar self-couplings~\cite{PakvasaSugawara1978}. The $S_3$ symmetry itself allows greater freedom, as illustrated by the fits to fermion masses and mixing obtained by Babu, Wu and Xu using general $S_3$-invariant Yukawa couplings in a three-Higgs-doublet model~\cite{BabuWuXu2024}. The original Clifford interaction imposes a further restriction when the charged leptons and down quarks share a Higgs vacuum. Their mass matrices are then proportional, giving identical mass ratios in the two sectors. 

We address these restrictions by introducing a second Yukawa interaction built from the internal charge operators. Subtracting the family average from each charge operator gives a family-dependent part from which a projector can be constructed. The projector acts on the fermion state from the left, while the Higgs operator acts from the right as before. This second contraction preserves the gauge and family symmetries and uses the same three Higgs doublets. Varying the relative strength of the two contractions independently in each sector allows a single real Higgs vacuum to reproduce the chosen charged-lepton and down-quark running masses. The fitted vacuum direction then determines the direction that the scalar potential must stabilise and enters the calculation of flavour change mediated by the additional neutral scalars.
\section{Algebraic setup}\label{Sec:Setup}

We begin by recalling the ingredients of the three-generation Clifford algebra model that are needed to investigate fermion masses. The fermion and Higgs representations were constructed in earlier work~\cite{Gresnigt2026Generations,Gresnigt2026Technical,Gresnigt2026Higgs}. Here we summarise the relevant states, family transformations and Yukawa maps, without repeating their explicit matrix construction. The distinction between these internal algebraic objects and the spacetime fields introduced below will be important when interpreting the resulting mass matrices.

\subsection{Fermion generations and internal Higgs operators}

The algebraic construction is formulated in the complex Clifford algebra $\mathcal A=\mathbb C\ell(10)\simeq\operatorname{Mat}_{32}(\mathbb C)$. The use of minimal ideals of $\mathbb C\ell(10)$ to represent one fermion generation with the full Standard Model gauge symmetry was developed in Ref.~\cite{Gresnigt2020StandardModel}. As in the earlier model~\cite{Gresnigt2026Technical}, we use five Witt pairs, which satisfy
\begin{equation}
\{a_i,a_j\}=0,\qquad \{a_i,a_j^\dagger\}=\delta_{ij}I_{32},\qquad i,j=1,\ldots,5.
\label{eq:setup-Witt}
\end{equation}
The Witt pairs allow us to construct primitive idempotents $f_{\sigma_1\cdots\sigma_5}=\prod_jp_j^{\sigma_j}$ from $p_j^+=a_ja_j^\dagger$ and $p_j^-=a_j^\dagger a_j$~\cite{Ablamowicz1995}. Their minimal left ideals $\mathcal Af_\sigma$ contain the selected weak-doublet and weak-singlet fermion subspaces. The Standard Model gauge generators act by commutators, reducing on these subspaces to left multiplication for colour and right multiplication for the weak action on doublets; the latter vanishes on singlets. Explicit matrix and gauge-generator conventions are given in Appendix~\ref{App:Scalar}.

We need only one set of fermion representatives to specify the family orbit. For the charged leptons, we take
\begin{equation}
L_1=a_4^\dagger f_{++++-},\qquad R_1=a_5^\dagger f_{+++-+}.
\label{eq:setup-lepton-seeds}
\end{equation}
Here $L_1$ and $R_1$ belong to the weak-doublet and weak-singlet sectors, respectively. Rather than introducing two further copies by hand, we obtain them through the $S_3$ family action inherited from the sedenion-motivated $\mathbb C\ell(8)$ construction~\cite{GourlayGresnigt2024}. Its $\dagger$-preserving generators $\psi$ and $\epsilon$ satisfy
\begin{equation}
\begin{gathered}
\psi^3=\epsilon^2=1,\qquad \epsilon\psi\epsilon=\psi^{-1},\\
L_a=\psi^{a-1}(L_1),\qquad R_a=\psi^{a-1}(R_1).
\end{gathered}
\label{eq:setup-family}
\end{equation}
The order-three transformation $\psi$ produces two additional, linearly independent generations. Because both automorphisms leave the gauge generators invariant, all three generations carry the same Standard Model quantum numbers despite having distinct algebraic representatives. Their matrix-unit forms and the corresponding down-quark states are given in Appendix~\ref{App:Mass}.

The Higgs operators connect the weak-doublet and weak-singlet sectors by right multiplication, $X\mapsto XH$~\cite{Gresnigt2026Higgs}. For a map from $\mathcal Af_A$ to $\mathcal Af_B$, they belong to $f_A\mathcal Af_B$. Since the primitive idempotents have rank one, this space is one-dimensional: the component map is fixed up to a coefficient once its source and target are specified. This does not, however, determine the relative coefficients of the complete spacetime Yukawa interaction.

In the conventions used here, the four first-generation Higgs components are
\begin{equation}
\begin{aligned}
H_d^0&=a_4a_5^\dagger f_{+++-+},& H_d^-&=a_4f_{+++-+},\\
H_u^+&=a_4f_{+++--},& H_u^0&=a_4a_5f_{+++--}.
\end{aligned}
\label{eq:setup-Higgs-seeds}
\end{equation}
For example, the neutral down-type component satisfies
\begin{equation*}
L_1H_d^0=R_1.
\end{equation*}
This shows explicitly how right multiplication connects the selected doublet and singlet states. It does not imply that every cross-generation product lies in the chosen right-handed span; the full Yukawa coefficients will be obtained from a trace pairing below.

The four operators form two electroweak doublets. With $Q_{\rm em}=T_3+Y/2$ and component phases chosen to give the standard weak generators, we write
\begin{equation}
\begin{gathered}
\boldsymbol H_d^{(1)}=\begin{pmatrix}H_d^0\\iH_d^-\end{pmatrix}\sim(1,2,-1),\\
\boldsymbol H_u^{(1)}=\begin{pmatrix}iH_u^+\\-H_u^0\end{pmatrix}\sim(1,2,+1),\\
\boldsymbol H_t^{(a)}=\psi^{a-1}\bigl(\boldsymbol H_t^{(1)}\bigr).
\end{gathered}
\label{eq:setup-internal-doublets}
\end{equation}
Here $(1,2,\pm1)$ denotes the $SU(3)_C\times SU(2)_L\times U(1)_Y$ representation. The family orbit contains twelve linearly independent internal Higgs components, which should not be confused with independent spacetime scalar fields. For the charged-lepton and down-quark masses we need only $H_a=\psi^{a-1}(H_d^0)$. The up-type operators complete the electroweak field content, but their Yukawa couplings are not fitted here. Details of the weak generators and component phases are given in Appendix~\ref{App:Scalar}.

The order-two generator also carries a significant sign. Let $F$ fix family 1 and exchange families 2 and 3. Then
\begin{equation}
\begin{gathered}
\epsilon(L_a)=-L_{F(a)},\qquad \epsilon(R_a)=R_{F(a)},\\
\epsilon\bigl(\boldsymbol H_t^{(a)}\bigr)=-\boldsymbol H_t^{(F(a))}.
\end{gathered}
\label{eq:setup-reflection}
\end{equation}
The signs distinguish the family representations: the left-handed lepton and Higgs orbits carry $\mathbf1'\oplus\mathbf2$, whereas the right-handed orbit carries $\mathbf1\oplus\mathbf2$~\cite{IshimoriEtAl2010}. In particular, the cyclically symmetric Higgs direction is a sign singlet, not a trivial singlet. This distinction will matter for the scalar potential in Sec.~\ref{Sec:Scalar}.

\subsection{Algebraic Yukawa couplings and generation overlaps}

The Higgs operators introduced above act on the internal fermion representatives. To construct the Yukawa interaction, we must associate a coefficient with each pair of left- and right-handed generation labels. Rather than introducing arbitrary family matrices, we obtain these coefficients from the overlaps of the images $L_rH_a$ with the right-handed states $R_s$. Following~\cite{Gresnigt2026Higgs}, we use the Hilbert--Schmidt inner product on $\mathcal A$,
\begin{equation}
\begin{gathered}
\langle X,W\rangle=\operatorname{Tr}(X^\dagger W)=32\langle X^\dagger W\rangle_0,\\
(Y_a)_{rs}=\operatorname{Tr}\bigl[(L_rH_a)^\dagger R_s\bigr],\qquad
Y(z)=\sum_a z_aY_a.
\end{gathered}
\label{eq:setup-trace}
\end{equation}
Here $\langle\cdot\rangle_0$ denotes the Clifford scalar part. Each matrix $Y_a$ is fixed by the chosen fermion representatives and Higgs map; the coefficients $z_a$ combine the three neutral Higgs directions and will be identified with spacetime vacuum expectation values below. Importantly, this construction does not require $L_rH_a$ to lie entirely in the span of the $R_s$. Components orthogonal to all the selected right-handed states do not contribute to the trace. The Yukawa tensors are thus defined by the pairings, not by requiring pointwise closure of the right action on the selected fermion space. Appendix~\ref{App:Mass} illustrates this distinction explicitly.

The overlaps between the fermion representatives themselves are equally important. Although the three generations are linearly independent, they are not orthogonal under the Hilbert--Schmidt inner product. Direct evaluation gives the same Gram matrix in the two chiral sectors,
\begin{equation}
\begin{gathered}
G_{rs}=\operatorname{Tr}(L_r^\dagger L_s)=\operatorname{Tr}(R_r^\dagger R_s),\\
G=\frac{3I_3+J}{4},\qquad J_{rs}=1.
\end{gathered}
\label{eq:setup-Gram}
\end{equation}
The diagonal entries are unity, while the off-diagonal entries are $1/4$. The eigenvalues of $G$ are $3/2$ in the family-symmetric direction and $3/4$ in the two orthogonal directions, so the Gram matrix is positive definite. These overlaps are a property of the algebraic family construction, not additional adjustable parameters.

The first-family Higgs map provides a useful illustration. Since $L_1H_1=R_1$, Eq.~\eqref{eq:setup-trace} gives $(Y_1)_{1s}=G_{1s}$, so the first row of $Y_1$ is $(1,1/4,1/4)$. The off-diagonal entries of a raw Yukawa tensor can therefore reflect the nonorthogonality of the family representatives rather than physical flavour mixing. In particular, the physical fermion masses cannot be determined from $Y(z)$ alone: they also depend on the kinetic terms of the associated spacetime fields. We now specify these terms and perform the corresponding canonical normalisation.

\subsection{Spacetime fields and canonical normalisation}

The preceding construction specifies internal fermion states and algebraic Yukawa coefficients. To interpret them as physical masses, we must also define a spacetime action. We associate independent spinor coefficient fields with the three generations, writing, for example, $\Psi_L(x)=\sum_a\chi_{L,a}(x)\otimes L_a$, and similarly for the right-handed states. Here the $L_a$ encode the internal representations, while the $\chi_{L,a}(x)$ are independent spacetime fields. Their kinetic terms are not fixed merely by the internal overlaps.

The same distinction arises for the Higgs fields. The three up-type and three down-type internal doublets introduced above need not give six independent spacetime doublets. An antilinear map relates the two internal orbits through electroweak doublet conjugation, while preserving the family transformation laws. Its construction is given in Appendix~\ref{App:Scalar}. We therefore choose three independent spacetime doublets $\Phi_a\sim(1,2,+1)$ with canonical scalar kinetic terms, assigning their internal coefficients by
\begin{equation}
\begin{gathered}
\mathcal L_{\rm kin}^{\Phi}=\sum_a(D_\mu\Phi_a)^\dagger D^\mu\Phi_a,\\
\varphi_{u,a}=\Phi_a,\qquad \varphi_{d,a}=\widetilde\Phi_a,\qquad
\widetilde\Phi_a=i\sigma_2\Phi_a^*.
\end{gathered}
\label{eq:setup-scalar-prescription}
\end{equation}
The coefficient doublets transform in the representations of their respective internal Higgs operators. Their duals enter the Yukawa pairing: $\widetilde\varphi_{d,a}=-\Phi_a$ and $\widetilde\varphi_{u,a}=\widetilde\Phi_a$. Absorbing the constant down-type sign into the sector coupling yields the usual contractions $\overline\ell_L\Phi_a e_R$, $\overline q_L\Phi_a d_R$, and up-type contractions with $\widetilde\Phi_a$. The family cycle permutes the $\Phi_a$, whereas the reflection sends $\Phi_a\mapsto-\Phi_{F(a)}$. Since the Hilbert--Schmidt pairing is conjugate-linear in its first argument, internal coefficients $\overline z_a$ yield the mass coefficients $z_aY_a$ of Eq.~\eqref{eq:setup-trace}, without further conjugation. These three doublets supply electroweak symmetry breaking in Sec.~\ref{Sec:Scalar}.

We next specify the fermion kinetic terms. Motivated by the inner product on the internal generation space, we choose $K_L=K_R=G$ for both charged-lepton and down-quark coefficient fields. With $z_a=\langle\Phi_a^0\rangle$ and coefficient columns $\chi_{L,R}$, the kinetic terms and original right-action interaction become
\begin{equation}
\begin{split}
\mathcal L_f={}&i\overline\chi_LG\gamma^\mu D_\mu\chi_L+i\overline\chi_RG\gamma^\mu D_\mu\chi_R\\
&-\bigl(\lambda_f\overline\chi_LY(z)\chi_R+\mathrm{h.c.}\bigr).
\end{split}
\label{eq:setup-fermion-action}
\end{equation}
Here $\lambda_f$ is a sector-wide dimensionless coupling. The use of $G$ is an explicit kinetic prescription, not a unique consequence of the algebra: the trace pairing fixes the $Y_a$, but does not determine $K_L$ and $K_R$. The choice of three canonically normalised scalar doublets is likewise an additional, symmetry-compatible assumption. Both prescriptions are held fixed below.

Since $G$ is positive definite, the field redefinitions $\chi_{L,R}^{\rm can}=G^{1/2}\chi_{L,R}$ bring the kinetic terms to canonical form. The corresponding Yukawa matrix is $G^{-1/2}Y(z)G^{-1/2}$, whose singular values, multiplied by $|\lambda_f|$, give the physical masses~\cite{KingPeddie2004,EspinosaIbarra2004,ChenEtAl2012,HornJohnson2012}. These are not, in general, the eigenvalues of the raw matrix $Y(z)$. Noncanonical kinetic structures can themselves contribute to hierarchies in flavour models~\cite{KakizakiYamaguchi2003,Kawamura2019}. Here, however, the metric is fixed and has condition number two. Our spectral analysis therefore concerns the Yukawa tensors with this prescribed normalisation.

It is convenient to work in an orthonormal family frame defined by
\begin{equation}
\begin{gathered}
Q=\begin{pmatrix}1/\sqrt3&1/\sqrt2&1/\sqrt6\\1/\sqrt3&-1/\sqrt2&1/\sqrt6\\1/\sqrt3&0&-2/\sqrt6\end{pmatrix},\\
D_G=\operatorname{diag}\!\left(\sqrt{\frac23},\frac2{\sqrt3},\frac2{\sqrt3}\right),\qquad Q^TGQ=D_G^{-2}.
\end{gathered}
\label{eq:setup-canonical-frame}
\end{equation}
The first column of $Q$ is the family-symmetric direction while the remaining columns span its orthogonal complement. The canonically normalised matrix, before its sector coupling, is $D_GQ^TY(z)QD_G$. The down-quark states give the same $G$ and $-Y_a$ in each colour, with vanishing cross-colour traces. Absorbing the common sign into the down-sector coupling allows both charged sectors to use the same canonical frame.

The cyclically averaged Yukawa tensor provides a useful illustration. Its raw and canonically normalised forms are
\begin{equation}
\begin{gathered}
\overline Y=\frac13\sum_aY_a=\frac38I_3+\frac18J=\frac12G,\\
G^{-1/2}\overline YG^{-1/2}=\frac12I_3.
\end{gathered}
\label{eq:setup-average}
\end{equation}
The raw matrix distinguishes the family-symmetric direction from a degenerate pair, whereas its canonically normalised form is proportional to the identity. Equivalently, equal vacuum coefficients $z_a=z_0$ give $D_GQ^TY(z)QD_G=(3z_0/2)I_3$, corresponding to three degenerate masses in either charged sector. Thus an eigenvalue hierarchy in the raw algebraic tensor does not automatically imply a physical mass hierarchy. We must instead examine the complete Higgs orbit, including unequal vacuum coefficients, to establish which spectra the original interaction permits.
\section{Fermion mass hierarchies}\label{Sec:Masses}

\subsection{Limitations of the original Yukawa interaction}

Section~\ref{Sec:Setup} showed that the cyclically symmetric Higgs direction yields three degenerate masses after canonical normalisation. We now investigate whether unequal Higgs expectation values can reproduce the observed fermion mass hierarchies using the original right-action interaction. We retain the fermion representatives, Hilbert--Schmidt pairing and kinetic prescription of Sec.~\ref{Sec:Setup}, so the physical masses are the singular values of the canonically normalised Yukawa matrix. One might expect additional Higgs directions to relax the restrictions of the original three-component construction. To distinguish a limitation of those directions from one of the interaction itself, we therefore allow the internal operator $H$ to range over the entire Clifford algebra $\mathcal A$, while retaining right multiplication, $X\mapsto XH$. This gives a more general test than varying the three original vacuum expectation values alone.

Despite the dimension of $\mathcal A$, only four independent complex combinations of $H$ enter the trace pairing between the selected fermion representatives. In the canonical family frame, the resulting matrices form the space
\begin{equation}
\mathcal R=\left\{\begin{pmatrix}d&p&q\\p&d&t\\q&-t&d\end{pmatrix}:d,p,q,t\in\mathbb C\right\}.
\label{eq:mass-right-image}
\end{equation}
The three original neutral Higgs directions span its $t=0$ subspace. Thus an arbitrary $H$ adds a fourth complex parameter to the matrix space. Remarkably, this additional freedom does not enlarge the attainable singular-value spectra. The restriction is therefore not an artefact of considering too few right-acting Higgs operators. Appendix~\ref{App:Mass} proves that both spaces give precisely the ordered masses satisfying
\begin{equation}
0\leq m_1\leq m_2\leq m_3,\qquad m_3\leq m_1+2m_2.
\label{eq:mass-right-bound}
\end{equation}
This condition is both necessary and sufficient, including for zero or degenerate masses. In particular, $m_1\leq m_2$ implies $m_3/m_2\leq3$ when $m_2>0$, whereas the tau-to-muon mass ratio is approximately $17$~\cite{XingZhangZhou2012}. The original interaction therefore cannot reproduce the charged-lepton hierarchy, regardless of the coefficients assigned to the three Higgs directions. More strongly, this remains true even when $H$ is unrestricted within $\mathcal A$.

The obstruction concerns this particular combination of fermion representatives, trace pairing, kinetic terms and right-action contraction; it is not imposed by $S_3$ symmetry alone~\cite{Gresnigt2026Higgs,KuboEtAl2003,GonzalezCanalesEtAl2013,BabuWuXu2024}. It consequently motivates the search for another symmetry-compatible contraction arising from the same algebraic framework.

\subsection{An additional algebraic Yukawa interaction}

We seek another interaction compatible with the gauge and family symmetries, but constructed from the same internal algebraic data rather than from arbitrary family Yukawa matrices. The charge operators already present in the three-generation Clifford construction suggest such an extension~\cite{Gresnigt2026Generations,Gresnigt2026Technical}. For the first family and its cyclic images, define
\begin{align}
Q_1&=\frac13\left(\sum_{j=1}^3a_ja_j^\dagger-3a_4a_4^\dagger\right),\qquad Q_a=\psi^{a-1}(Q_1),\nonumber\\
\overline Q&=\frac13\sum_{a=1}^3Q_a,\qquad Z_a=Q_a-\overline Q,\qquad A_a=\frac{I_{32}}2+Z_a.
\label{eq:mass-projectors}
\end{align}
The operators $Q_a$ form a family orbit, whereas their average $\overline Q$ is invariant. Subtracting that average removes the family-independent contribution and isolates the part $Z_a$ that distinguishes the three positions in the orbit. In the chosen representation, $Z_a$ is Hermitian and satisfies $Z_a^2=I_{32}/4$. Thus $A_a$ projects onto its $+1/2$ eigenspace, with $A_a^\dagger=A_a^2=A_a$ and $\operatorname{rank}A_a=16$. The choice of $A_a$ is thus motivated by an existing spectral decomposition within the Clifford representation. In particular, no freely adjustable $3\times3$ family matrix is introduced at this stage.

The $A_a$ act on the full $32$-dimensional internal space, not on individual fermion generations: $\sum_aA_a=3I_{32}/2$, and they are not mutually orthogonal family projectors. Replacing them by $I_{32}-A_a$ yields the difference between the original and new contractions, rather than a third independent tensor. This complementarity explains why the two-contraction space considered below is unaffected by choosing the opposite charge eigenspaces. It does not establish that the present contraction is unique, and other algebraically compatible extensions may exist.

We must next establish that inserting $A_a$ preserves the relevant symmetries. Since each $A_a$ commutes with the Standard Model gauge generators, for any such generator $B$ we have
\begin{equation}
[B,A_aXH_a]=A_a[B,X]H_a+A_aX[B,H_a].
\label{eq:mass-covariance}
\end{equation}
The first term on the right-hand side describes the gauge action on the fermion state, and the second that on the Higgs operator. The projector contributes no additional gauge transformation. Under $S_3$, the $A_a$ are permuted without the reflection sign carried by $H_a$, so the projected maps transform in the same way as the original right-action maps. Both sets can therefore be contracted with the same dual scalar multiplet introduced in Sec.~\ref{Sec:Setup}. The new interaction preserves the gauge and family symmetries without requiring additional Higgs fields.

Using the Hilbert--Schmidt pairing of Sec.~\ref{Sec:Setup}, the projected maps give
\begin{equation}
\begin{gathered}
A_aL_rH_a=G_{ar}R_a,\\
(T_a)_{rs}=\operatorname{Tr}\!\left[(A_aL_rH_a)^\dagger R_s\right]=G_{ra}G_{as}.
\end{gathered}
\label{eq:mass-new-tensor}
\end{equation}
The first identity shows that the projector-dressed map takes the selected left-handed representatives into the corresponding right-handed direction, with coefficients fixed by their overlaps. Taking the trace against $R_s$ then gives the outer product of the $a$th Gram-matrix column and row. Hence each $T_a$ has rank one, and the vacuum-weighted combination is $\sum_a z_aT_a=G\operatorname{diag}(z)G$. This is a direct consequence of the internal representatives and their Hilbert--Schmidt pairing, rather than an imposed form for the Yukawa tensors. Appendix~\ref{App:Mass} gives the explicit matrix verification and family transformation laws.

For comparison with the original interaction, introduce the following coordinates in the orthonormal family frame:
\begin{equation}
\begin{gathered}
c=\frac{z_1+z_2+z_3}{4},\qquad p=\frac{\sqrt3(z_1-z_2)}4,\\
q=\frac{z_1+z_2-2z_3}{4}.
\end{gathered}
\label{eq:mass-vacuum-coordinates}
\end{equation}
Here $c$ describes the cyclically symmetric direction, which belongs to the sign representation under reflection as explained in Sec.~\ref{Sec:Setup}; $p$ and $q$ span the two independent departures from that direction. The change of coordinates is invertible and therefore imposes no restriction on the vacuum. Its advantage is that it makes the structure of the canonically normalised matrices transparent. After canonical normalisation, the original and projector-dressed interactions give
\begin{equation}
\begin{gathered}
O(z)=\begin{pmatrix}2c&p&q\\p&2c&0\\q&0&2c\end{pmatrix},\\
N(z)=\begin{pmatrix}2c&p&q\\p&c+q/2&p/2\\q&p/2&c-q/2\end{pmatrix}.
\end{gathered}
\label{eq:mass-two-matrices}
\end{equation}
Unlike the equal-diagonal arrow matrix $O(z)$, the new matrix $N(z)$ has a different lower $2\times2$ block, including a generally nonzero $23$ entry. This changes the allowed family dependence of the Yukawa interaction while keeping the fermion representatives, Gram matrix and scalar multiplet fixed. The two matrices remain correlated through the same three vacuum coefficients, rather than constituting independently adjustable Yukawa matrices.

Although each coefficient $N_a$ of $z_a$ in $N(z)$ has rank one, their singular vectors need not be aligned. Their relative orientations are fixed by the algebra, and rank one does not imply the singular alignment considered in~\cite{RodejohannSaldanaSalazar2019}. The resulting neutral-scalar flavour effects are examined in Sec.~\ref{Sec:Flavour}.

Writing $O_a=\partial O/\partial z_a$ and $N_a=\partial N/\partial z_a$, both contractions contribute to the charged-lepton and down-quark mass matrices:
\begin{equation}
\begin{gathered}
M_f(z)=\lambda_fO(z)+\mu_fN(z)=\sum_a z_a\mathcal Y_{fa},\\
\mathcal Y_{fa}=\lambda_fO_a+\mu_fN_a,\qquad f=e,d.
\end{gathered}
\label{eq:mass-pencil}
\end{equation}
The coefficients $\lambda_f$ and $\mu_f$ are dimensionless sector-wide couplings associated with the two contractions. They can differ between the charged-lepton and down-quark representations, but within a given sector they multiply algebraically fixed tensors, not arbitrary entries of a family Yukawa matrix. The down-quark tensors differ by an overall sign in each colour, absorbed into these couplings; their colour structure is diagonal and their kinetic matrix remains $G$.

An essential restriction is that both contractions use the same vacuum coefficients $z_a$, and the charged-lepton and down-quark sectors share that vacuum as well. Assigning separate directions to the two contractions or to the two sectors would introduce additional freedom and define a different model. For canonically normalised spacetime fermions, the gauge-invariant interaction is
\begin{equation}
\begin{split}
\mathcal L_Y^{e,d}=-\sum_a\bigl[&\overline\ell_L\mathcal Y_{ea}\Phi_a e_R\\
&+\overline q_L\mathcal Y_{da}\Phi_a d_R+\mathrm{h.c.}\bigr].
\end{split}
\label{eq:mass-Yukawa-action}
\end{equation}
We can now determine whether this two-contraction model admits the hierarchies excluded in the preceding subsection.

\subsection{Hierarchical mass spectra}

Consider first one charged fermion sector, varying $z_a$ at fixed couplings $(\lambda,\mu)$. We ask when the singular values of $M(z)=\lambda O(z)+\mu N(z)$ can develop arbitrarily large ratios. Rank-one limits are particularly relevant here. A nonzero rank-one matrix has one nonzero singular value and two vanishing singular values, so it represents the limiting configuration in which one generation remains massive while the other two become massless. If a full-rank family approaches such a matrix, $m_3/m_2$ diverges. Conversely, once the vacuum coordinates are normalised to remove an overall scale, they form a compact parameter space. The associated mass-matrix map is injective for every nonzero coupling pair, as shown in Appendix~\ref{App:Mass}; thus an unbounded $m_3/m_2$ requires a nonzero rank-one limit. This reduces the hierarchy question to a classification of such matrices.

For $\mu\ne0$, a nonzero rank-one matrix occurs precisely when
\begin{equation}
\frac\lambda\mu\in\left\{0,-\frac12,-1,-\frac32\right\}.
\label{eq:mass-rank-one-ratios}
\end{equation}
This classification holds for complex couplings and vacua. Appendix~\ref{App:Mass} gives the proof. The pure additional interaction, $\lambda=0$, provides a simple example: each coefficient $N_a$ has rank one, as established above. The remaining special ratios result from combining the two algebraically fixed contractions. Nevertheless, a rank-one member alone is insufficient for three nonzero masses: full-rank matrices must also exist nearby. At $\lambda/\mu=-1$, however, every matrix in the family is singular, so no choice of vacuum produces three nonzero masses. At each of the other three listed ratios, $m_3/m_2$ grows without bound as full-rank matrices approach rank one. For every other fixed coupling pair that admits full-rank matrices, including the pure original interaction $\mu=0$, this ratio is bounded. This statement distinguishes unbounded hierarchies from finite ones: a finite upper bound at a generic coupling ratio need not be small enough to exclude a given observed mass ratio.

The absence of a bound on $m_3/m_2$ is not sufficient by itself to produce a hierarchy between all three generations. In particular, the two lighter masses might remain comparable as they approach zero. We therefore ask whether both successive ratios, $m_3/m_2$ and $m_2/m_1$, can become arbitrarily large. An explicit construction shows that this is possible while retaining both contractions. Choose $(\lambda,\mu)=(-3/2,1)$, write $z=v_*\widehat z$ with $v_*>0$, and take
\begin{equation}
\widehat z=4(J-2I)\begin{pmatrix}\varepsilon^2\\\varepsilon\\1\end{pmatrix},\qquad 0<\varepsilon\ll1.
\label{eq:mass-hierarchy-path}
\end{equation}
Here $J$ is the all-ones matrix defined in Sec.~\ref{Sec:Setup}. For every $\varepsilon>0$ the resulting matrix is of full rank. As $\varepsilon\to0$, its singular values behave as
\begin{equation}
\begin{gathered}
\frac{(m_1,m_2,m_3)}{v_*}\sim\left(\frac92\varepsilon^2,\frac{24}{5}\varepsilon,5\right),\\
\det\!\left(\frac{M}{v_*}\right)=-108\varepsilon^3.
\end{gathered}
\label{eq:mass-hierarchy-asymptotics}
\end{equation}
As derived in Appendix~\ref{App:Mass}, the heaviest mass approaches a nonzero value, while the other two scale as $\varepsilon$ and $\varepsilon^2$. Indeed, $m_3/m_2\sim25/(24\varepsilon)$ and $m_2/m_1\sim16/(15\varepsilon)$, so both successive ratios diverge as $\varepsilon\to0$. The resulting spectra therefore exhibit a genuine three-level hierarchy, not simply one heavy state separated from two comparably light states. Multi-doublet vacuum structures have also been used to obtain mass hierarchies~\cite{BaekEtAl2024}, but here the two tensors are fixed by the Clifford construction. This establishes that the two specific Clifford-derived contractions permit such spectra even when the original vertex is retained. It is an algebraic existence statement: the special coupling ratio and vacuum path are inputs, and their dynamical origin remains to be explained.

We must also address a separate limitation of the original interaction. If the charged-lepton and down-quark sectors share one vacuum, their mass matrices with only $O(z)$ differ by an overall sector coupling and are necessarily proportional. That is incompatible with treating their markedly different mass patterns as independent constraints. The additional term removes this enforced proportionality: $O(z)$ and $N(z)$ are linearly independent for every nonzero $z$, and their three-dimensional matrix spaces intersect only at zero (Appendix~\ref{App:Mass}). Thus, for $z\ne0$ and $\eta\ne0$,
\begin{equation}
M_e(z)=\eta M_d(z)\quad\Longleftrightarrow\quad(\lambda_e,\mu_e)=\eta(\lambda_d,\mu_d).
\label{eq:mass-proportionality}
\end{equation}
Nonproportional coupling pairs therefore remove enforced matrix proportionality. This statement concerns matrices, not their singular values: two nonproportional matrices may still have proportional singular spectra. Consequently, removing the algebraic restriction does not yet demonstrate that both observed mass patterns can be fitted using one and the same vacuum.

\subsection{Charged-lepton and down-quark masses from a common vacuum}

The results of the preceding subsection show that large hierarchies are permitted and that a shared vacuum no longer forces proportional mass matrices. They do not, however, prove that the two observed spectra can be reproduced simultaneously. We now test this stronger requirement with a single real vacuum $z_a$ and independent sector-wide couplings for the charged-lepton and down-quark sectors. As a first benchmark we take the Standard Model $\overline{\mathrm{MS}}$ running masses at $M_Z$ quoted in Ref.~\cite{XingZhangZhou2012}, in MeV:
\begin{equation}
\begin{gathered}
(m_e,m_\mu,m_\tau)=(0.486570154,102.7181337,1746.17),\\
(m_d,m_s,m_b)=(2.82,57,2860).
\end{gathered}
\label{eq:mass-inputs}
\end{equation}
To separate mass ratios from overall scales, set $z=v_*\widehat z$, with $v_*>0$, and choose $c(\widehat z)=1$. The condition $c(\widehat z)=1$ fixes an otherwise redundant overall rescaling of the vacuum direction; it does not yet specify the physical electroweak scale, which is imposed in Sec.~\ref{Sec:Scalar}. Defining $r_f=\lambda_f/\mu_f$ and $\widehat\mu_f=v_*\mu_f$, we have $M_f=\widehat\mu_f F_{r_f}$, where $F_r=rO(\widehat z)+N(\widehat z)$. We restrict attention to the real branch needed for the scalar analysis below. The four shape parameters $(p,q,r_e,r_d)$ determine the four mass ratios, while $(\widehat\mu_e,\widehat\mu_d)$ fix the two overall scales. The parameters $p$ and $q$ describe the same vacuum in both sectors; choosing them independently would evade the restriction we wish to test. Although the number of parameters matches the number of targets, that counting does not guarantee a solution within this particular matrix family.

For the inputs above, one solution is represented by
\begin{equation}
\begin{gathered}
\begin{aligned}
p&\simeq0.0223930,&q&\simeq-1.99568,\\
r_e&\simeq0.130486,&r_d&\simeq0.0364469,
\end{aligned}\\
\widehat z\simeq(0.0287372,-0.0229772,3.99424).
\end{gathered}
\label{eq:mass-benchmark-shape}
\end{equation}
with $\widehat\mu_e\simeq386.663\,\mathrm{MeV}$ and $\widehat\mu_d\simeq690.780\,\mathrm{MeV}$. At the certified solution, the singular values of the two mass matrices reproduce all six target masses. On this real branch the charged-lepton eigenvalues are positive, whereas the light down-quark eigenvalue is negative and its physical mass is the absolute value. Its right-handed sign transformation must be retained when calculating the neutral-scalar couplings in Sec.~\ref{Sec:Flavour}.

The displayed numbers alone would amount only to a numerical fit. To establish the solution more firmly, Appendix~\ref{App:Mass} expresses the exact rational target ratios as four polynomial equations and provides a rational interval certificate. This proves that an exact real solution exists and is unique within a specified small parameter box, without relying on diagonalisation of rounded matrices. A second, uniform certificate demonstrates a finite neighbourhood of feasible targets. The two lepton ratios may vary independently by relative amounts up to $10^{-8}$ and the two down-quark ratios by up to $1\%$, with the heavy masses held fixed, and every target in this neighbourhood has a nearby common-vacuum solution. The mass-ratio Jacobian is nonsingular there. The result establishes local existence and independence of the four ratios; it neither proves global uniqueness nor assigns experimental confidence to the stated intervals.

This benchmark demonstrates compatibility rather than a parameter-free mass prediction. The four ratios are fitted by four shape parameters, and two more parameters set the mass scales. The vacuum direction and coupling ratios have not been derived dynamically.

For comparison, we repeat the fit using the more recent Standard Model $\overline{\mathrm{MS}}$ Yukawa central values and running vacuum expectation value $\bar v(M_Z)$ of Ref.~\cite{AntuschHinzeSaad2026}. The comparison targets are defined by $m_f^{\mathrm{cmp}}=y_f\bar v/\sqrt2$ from the rounded central quantities quoted in that reference. The running $\bar v(M_Z)$ used to define these inputs must be distinguished from the electroweak normalisation based on $G_F$ that we adopt for our scalar fields in Sec.~\ref{Sec:Scalar}. The same real branch gives
\begin{equation}
\begin{gathered}
\begin{aligned}
p&\simeq0.0210952,&q&\simeq-1.996007,\\
r_e&\simeq0.130873,&r_d&\simeq0.0342770,
\end{aligned}\\
(\widehat\mu_e,\widehat\mu_d)\simeq(386.3513,692.8928)\,\mathrm{MeV}.
\end{gathered}
\label{eq:mass-updated-benchmark}
\end{equation}
Appendix~\ref{App:Mass} also supplies an independent rational certificate and a finite feasibility neighbourhood for this second set of central inputs. The electroweak normalisation and the other independent inputs to the scalar and flavour calculations are held fixed in this comparison. The agreement at two input sets supports the compatibility of the common-vacuum construction but does not determine its dynamical origin. The remaining question is whether the fitted vacuum direction can be realised as a stable minimum of a scalar potential.

\section{Vacuum and scalar spectrum}\label{Sec:Scalar}

The common real vacuum found in Sec.~\ref{Sec:Masses} must also arise as a stable minimum of a scalar potential. We assume that the three canonically normalised doublets of Sec.~\ref{Sec:Setup} account for all electroweak symmetry breaking. The Fermi constant then fixes the overall scale left free by the mass-ratio fit. Writing
\begin{equation}
\begin{gathered}
z_a=\langle\Phi_a^0\rangle=\frac{v_a}{\sqrt2}=v_*\widehat z_a,\qquad v^2=\sum_a v_a^2=(\sqrt2G_F)^{-1},\\
n=\frac{\widehat z}{\|\widehat z\|},\qquad v_* =\frac{v}{\sqrt2\|\widehat z\|}.
\end{gathered}
\label{eq:scalar-ew-budget}
\end{equation}
and using $G_F=1.1663787\times10^{-5}\,\mathrm{GeV}^{-2}$, the certified solution gives
\begin{equation}
\begin{gathered}
v\simeq246.21965\,\mathrm{GeV},\qquad v_*\simeq43.58681\,\mathrm{GeV},\\
n\simeq(0.00719436,-0.00575235,0.99995757)^T.
\end{gathered}
\label{eq:scalar-vacuum}
\end{equation}
The dimensionful fit parameters $\widehat\mu_f$ now determine the Yukawa couplings $\mu_f=\widehat\mu_f/v_*$ and $\lambda_f=r_f\mu_f$:
\begin{equation}
\begin{aligned}
(\lambda_e,\mu_e)&\simeq(0.00115755,0.00887109),\\
(\lambda_d,\mu_d)&\simeq(0.000577623,0.0158484).
\end{aligned}
\label{eq:scalar-yukawa-couplings}
\end{equation}

\subsection{The obstruction with exact $S_3$ symmetry}

Kaneko et al. studied how the Higgs vacuum determines fermion mass textures in models with permutation flavour symmetry~\cite{kaneko2007flavor}. We first ask whether this direction can be an isolated minimum with exact $S_3$ symmetry. The general CP-conserving three-doublet potential is known~\cite{DasDey2014,EmmanuelCostaEtAl2016}. Although our family action is $\operatorname{sgn}(g)P(g)$ rather than $P(g)$, the common sign cancels in every bilinear $\Phi_a^\dagger\Phi_b$, leaving its quadratic and quartic invariants unchanged. In the orthonormal family frame,
\begin{equation}
\begin{gathered}
(S,D_1,D_2)^T=Q^T(\Phi_1,\Phi_2,\Phi_3)^T,\\
\langle(S^0,D_1^0,D_2^0)\rangle=\frac{(s,x,y)}{\sqrt2}.
\end{gathered}
\label{eq:scalar-family-frame}
\end{equation}
$S$ is a pseudosinglet and $(D_1,D_2)$ a family doublet. For a real neutral configuration, the only family-angle-dependent term is $\kappa_4sy(3x^2-y^2)/2$; the full potential is given in Appendix~\ref{App:Scalar}. Taking the angular combination of the stationarity equations yields
\begin{equation}
(y\partial_x-x\partial_y)V_0=\frac{3\kappa_4}{2}sx(3y^2-x^2)=0.
\label{eq:scalar-angular-obstruction}
\end{equation}
For the certified vacuum, $s\ne0$, $x\ne0$ and $x^2\ne3y^2$ throughout the certificate box. Stationarity therefore forces $\kappa_4=0$. In this limit the full potential acquires an accidental $SO(2)$ symmetry rotating $(D_1,D_2)$~\cite{EmmanuelCostaEtAl2016}. The fitted vacuum breaks this continuous symmetry, yielding a physical massless CP-even scalar. It cannot therefore be an isolated tree-level minimum of the exact-$S_3$ potential. This obstruction applies to the specified real direction, not to all $S_3$ vacua. Analogous accidental symmetries and their lifting by soft terms occur in complex-vacuum models~\cite{KuncinasEtAl2020}.

\subsection{A softly broken potential with the required vacuum}

We therefore allow dimension-two soft breaking while leaving the Yukawa tensors unchanged. Such terms have been used to obtain heavy additional scalars in $S_3$ models~\cite{KuboOkadaSakamaki2004}, although their effect on vacuum alignment depends on the quadratic matrix~\cite{VarzielasIvanovLevy2021}. We choose this matrix to annihilate the fitted unit direction $n$ and consider
\begin{equation}
\begin{gathered}
V=\lambda_V\left(\sum_a\Phi_a^\dagger\Phi_a-\frac{v^2}{2}\right)^2+\sum_{a,b}\Phi_a^\dagger\mathsf S_{ab}\Phi_b,\\
\mathsf S=M^2(I-nn^T),\qquad \lambda_V>0,\quad M>0.
\end{gathered}
\label{eq:scalar-soft-potential}
\end{equation}
The radial quartic has $U(3)$ family symmetry, as in maximally symmetric three-doublet models~\cite{DarvishiMasouminiaPilaftsis2021}; only the quadratic term breaks $S_3$. Both terms are nonnegative throughout the complex field space, including charged configurations. Since $\mathsf S$ is proportional to the projector onto the subspace orthogonal to $n$, its contribution vanishes precisely when each weak component has family dependence proportional to $n$. The radial term fixes the overall magnitude, so both terms vanish exactly when
\begin{equation}
\Phi_a=n_aH,\qquad H^\dagger H=\frac{v^2}{2}.
\label{eq:scalar-global-minima}
\end{equation}
Every such configuration is gauge equivalent to $H=(0,v/\sqrt2)^T$. The fitted vacuum is thus the unique gauge orbit of global minima, not merely a stationary point of the neutral potential. More generally, the construction works for any positive semidefinite quadratic matrix with kernel $\mathbb Cn$ and two strictly positive remaining eigenvalues.

\subsection{Scalar spectrum and alignment}

The spectrum is simplest in the Higgs basis. Let $R$ be real orthogonal with first row $n^T$, and define $\mathcal H_k=\sum_aR_{ka}\Phi_a$; an explicit $R$ appears in Appendix~\ref{App:Scalar}. Only $\mathcal H_1$ has a vacuum expectation value, and we write
\begin{equation}
\begin{gathered}
\mathcal H_1=\begin{pmatrix}G^+\\(v+h+iG^0)/\sqrt2\end{pmatrix},\\
\mathcal H_k=\begin{pmatrix}H_k^+\\(H_k+iA_k)/\sqrt2\end{pmatrix},\qquad k=2,3.
\end{gathered}
\label{eq:scalar-higgs-basis}
\end{equation}
In the original family basis, the quadratic mass matrices are
\begin{equation}
\mathcal M_{\mathrm{even}}^2=\mathsf S+2\lambda_Vv^2nn^T,\qquad \mathcal M_{\mathrm{odd}}^2=\mathcal M_\pm^2=\mathsf S,
\label{eq:scalar-hessians}
\end{equation}
giving
\begin{equation}
\begin{gathered}
m_h^2=2\lambda_Vv^2,\\
m_{H_k}^2=m_{A_k}^2=m_{H_k^\pm}^2=M^2,\qquad k=2,3.
\end{gathered}
\label{eq:scalar-spectrum}
\end{equation}
Of the twelve real field components, three are gauge Goldstone modes. The nine physical degrees of freedom are $h$, four further neutral scalars and two complex charged scalars. Their mass degeneracies at $M$ reflect the particular potential chosen, rather than a necessary prediction of the Clifford construction.

The field $h$ lies along the vacuum direction and has Standard Model tree-level couplings to $W$ and $Z$. Its charged-lepton and down-quark Yukawa couplings are also mass proportional, since the same vacuum-weighted Yukawa combinations define their mass matrices. This scalar alignment, discussed more generally in $S_3$ three-doublet models~\cite{GomezBockEtAl2021}, does not align the Yukawa matrices of the other neutral scalars in flavour space.

With $m_h=125\,\mathrm{GeV}$, one obtains $\lambda_V\simeq0.12886786$. The heavy scalar masses can be increased through $M$ without enlarging the quartic coupling, illustrating soft-breaking decoupling~\cite{CarroloEtAl2021}. This potential establishes the existence of a stable completion but does not explain the fitted direction $n$, which is an input to $\mathsf S$. The special quartic relations and a large hierarchy between $M$ and $v$ also require radiative analysis. We next consider the neutral-scalar flavour effects of this completion.

\section{Neutral-scalar flavour violation}\label{Sec:Flavour}

The vacuum-aligned light Higgs of Sec.~\ref{Sec:Scalar} has mass-proportional Yukawa couplings, but the Yukawa matrices of the two orthogonal doublets need not be diagonal in the fermion mass basis~\cite{BrancoEtAl2012}. Natural flavour conservation~\cite{GlashowWeinberg1977,Paschos1977}, proportional alignment~\cite{PichTuzon2009}, singular alignment~\cite{RodejohannSaldanaSalazar2019} and residual family symmetries~\cite{VarzielasTalbert2020} provide ways of avoiding or restricting tree-level neutral flavour change. We ask first whether the Clifford-derived tensors admit a common mass basis, and then whether their flavour-changing couplings satisfy phenomenological constraints.

\subsection{Yukawa misalignment and neutral-scalar couplings}

For the three canonically normalised matrices $\mathcal Y_{fa}$ in Eq.~\eqref{eq:mass-pencil}, no nonzero complex pair $(\lambda_f,\mu_f)$ admits independent unitaries $U_L,U_R$ for which all $U_L^\dagger\mathcal Y_{fa}U_R$ are diagonal. This includes either contraction separately and every rank-one branch of Sec.~\ref{Sec:Masses}. Appendix~\ref{App:Scalar} proves the result using the necessary normality and commutativity conditions on $\mathcal Y_{fa}\mathcal Y_{fb}^\dagger$ for the full complex coupling space. The obstruction is independent of the vacuum and invariant under scalar-basis changes. At the benchmark, where the fermion masses are nondegenerate, it implies physical off-diagonal neutral couplings, although it does not fix their magnitude.

Let $R$ be the real Higgs-basis rotation with first row $n^T$, so that $\mathcal H_\alpha=\sum_aR_{\alpha a}\Phi_a$ and only $\mathcal H_1$ has a vacuum expectation value. In the positive-mass basis, define
\begin{equation}
\begin{gathered}
U_{Lf}^\dagger M_fU_{Rf}=D_f=\operatorname{diag}(m_{1f},m_{2f},m_{3f})>0,\\
\Gamma_\alpha^f=U_{Lf}^\dagger\!\left(\sum_aR_{\alpha a}\mathcal Y_{fa}\right)U_{Rf},
\qquad \Gamma_1^f=\frac{\sqrt2D_f}{v}.
\end{gathered}
\label{eq:flavour-mass-basis}
\end{equation}
The light-Higgs matrix $\Gamma_1^f$ is mass proportional, while $\Gamma_{2,3}^f$ govern the additional neutral scalars. At the real benchmark, a real orthogonal matrix $O_f$ diagonalises $M_f$ with signed eigenvalues. Positive masses require $U_{Lf}=O_f$ and $U_{Rf}=O_fP_f$, with $P_e=I$ and $P_d=\operatorname{diag}(-1,1,1)$. The down-sector sign multiplies the corresponding column of every $\Gamma_\alpha^d$; simply taking the absolute value of the negative eigenvalue would give incorrect physical couplings.

\subsection{Neutral-scalar exchange and meson mixing}

The additional neutral interactions in each fermion sector are
\begin{equation}
\begin{split}
\mathcal L_{\mathrm n}^f={}&-\sum_{\alpha=2,3}\frac{H_\alpha}{\sqrt2}\bigl(\bar f_L\Gamma_\alpha^ff_R+\mathrm{h.c.}\bigr)\\
&-\sum_{\alpha=2,3}\frac{iA_\alpha}{\sqrt2}\bigl(\bar f_L\Gamma_\alpha^ff_R-\mathrm{h.c.}\bigr).
\end{split}
\label{eq:flavour-neutral-interaction}
\end{equation}
Both CP-even and CP-odd fields must be included. For the degenerate spectrum of Sec.~\ref{Sec:Scalar}, their relative signs cancel equal-chirality $\Delta F=2$ contributions but add in the opposite-chirality channel. With colour contracted within each quark bilinear, the effective-Hamiltonian coefficient is
\begin{equation}
\begin{gathered}
Q_4^{ij}=(\bar f_iP_Lf_j)(\bar f_iP_Rf_j),\qquad P_{L,R}=\frac{1\mp\gamma_5}{2},\\
W_4^{ij}(M)=-\frac{\kappa_{ij}(M)}{M^2},\qquad
\kappa_{ij}=\sum_{\alpha=2,3}(\Gamma_\alpha)_{ij}(\Gamma_\alpha)_{ji}^*.
\end{gathered}
\label{eq:flavour-lr-matching}
\end{equation}
The effective-Lagrangian coefficient has the opposite sign. Appendix~\ref{App:Scalar} gives the nondegenerate matching, for which the cancellation need not hold. The full degenerate exchange is invariant under complex $U(2)$ rotations of the two zero-VEV doublets, although the individual CP-even and CP-odd terms need not be. At $M_Z$ the benchmark gives
\begin{equation}
(\kappa_{ds},\kappa_{db},\kappa_{sb})=(-7.76632,-1.72510,1.40115)\times10^{-5}.
\label{eq:flavour-kappa-values}
\end{equation}

The connection between fermion hierarchies and neutral-scalar flavour constraints has been examined in other three-doublet constructions~\cite{BabuWuXu2024,AltmannshoferToner2025}. We assess the down-sector benchmark by requiring the absolute new-physics contribution to each neutral-meson mass splitting to satisfy
\begin{equation}
\mathcal R_P\equiv\frac{|2M_{12}^{P,\mathrm{new}}|}{\Delta M_P^{\mathrm{exp}}}<0.1,\qquad P=K,B_d,B_s.
\label{eq:flavour-mixing-criterion}
\end{equation}
Here ten percent is a chosen allowance, not an experimental confidence bound, and no cancellation with the Standard Model amplitude is assumed. We evolve the down Yukawa matrices from $M_Z$ to $M$ in leading QCD, match the tree-level exchange at $M$, and evolve the coefficients to $\mu_E=160\,\mathrm{GeV}$ in the BMU basis~\cite{BurasJagerUrban2001}. Lower-scale evolution and hadronic matrix elements enter through the $P$ factors of Ref.~\cite{AebischerEtAl2020Atlas}. At this order, squared Yukawa running times scalar-LR Wilson running above $\mu_E$ is independent of $M$; the combined factor from the $M_Z$ Yukawas to the interface is $0.9219213271$. Conventions and numerical inputs are detailed in Appendix~\ref{App:Scalar}.

At the finite benchmark scale $M=2000\,\mathrm{TeV}$, we obtain
\begin{equation}
\begin{aligned}
\mathcal R_K&=2.53104\times10^{-3},\\
\mathcal R_{B_d}&=5.06943\times10^{-6},\\
\mathcal R_{B_s}&=1.68574\times10^{-7}.
\end{aligned}
\label{eq:flavour-finite-meson-values}
\end{equation}
Kaon mixing is the most restrictive channel and, within this leading-QCD prescription and ten-percent criterion, requires $M>318.2\,\mathrm{TeV}$. The $2000\,\mathrm{TeV}$ benchmark also passes the additional amplitude tests of Appendix~\ref{App:Scalar}, which include stated allowances for hadronic and short-distance corrections. These bounds apply to the specified benchmark and scalar spectrum, not to unexamined flavour sectors.

\subsection{Charged-lepton flavour violation and phenomenological limitations}

The same neutral scalars mediate $\mu\to3e$. In the massless-electron contact approximation,
\begin{equation}
\begin{gathered}
a=\frac1{M^2}\sum_{\alpha=2,3}(\Gamma_\alpha^e)_{e\mu}(\Gamma_\alpha^e)_{ee}^*,\\
b=\frac1{M^2}\sum_{\alpha=2,3}(\Gamma_\alpha^e)_{\mu e}^*(\Gamma_\alpha^e)_{ee},\\
\operatorname{Br}(\mu\to3e)=\frac{|a|^2+|b|^2}{32G_F^2}.
\end{gathered}
\label{eq:flavour-mu3e}
\end{equation}
The normalisation follows Ref.~\cite{KunoOkada2001}, with a direct phase-space derivation in Appendix~\ref{App:Scalar}. At $M=2000\,\mathrm{TeV}$ we obtain $\operatorname{Br}(\mu\to3e)=8.17\times10^{-28}$, well below the published SINDRUM bound of $10^{-12}$ at $90\%$ confidence~\cite{BellgardtEtAl1988}. The cancellation-free contact test in Appendix~\ref{App:Scalar} gives the same conclusion. For the present benchmark, neutral-meson mixing is therefore much more restrictive than this charged-lepton decay.

Repeating the analysis with the updated mass benchmark of Eq.~\eqref{eq:mass-updated-benchmark}, at the same $M=2000\,\mathrm{TeV}$ and with the same QCD and hadronic inputs, gives $\mathcal R_K=2.53990\times10^{-3}$, $\mathcal R_{B_d}=5.09972\times10^{-6}$ and $\mathcal R_{B_s}=1.68344\times10^{-7}$, together with $\operatorname{Br}(\mu\to3e)=7.26\times10^{-28}$. The kaon-dominated lower scale becomes $318.7\,\mathrm{TeV}$ under the same leading-QCD ten-percent criterion, and the conditional amplitude tests in Appendix~\ref{App:Scalar} remain satisfied.

Thus the fixed Yukawa tensors necessarily give neutral flavour-changing couplings, but the explicit heavy-scalar completion satisfies the specified tests. This does not constitute a complete flavour analysis. Up-quark Yukawas and the physical CKM phase are still needed to examine charged-scalar, CP-violating and dipole observables. Electroweak and Yukawa evolution, radiative scalar splittings and quantum vacuum stability also remain open; in particular, the equal-chirality cancellation relies on the tree-level scalar degeneracies of Sec.~\ref{Sec:Scalar}.

\section{Discussion}\label{Sec:Discussion}

With the internal Gram matrix chosen as the fermion kinetic metric, the original right-action Yukawa interaction excludes the observed charged-lepton and down-quark mass hierarchies. Allowing the right-acting operator to range over the full Clifford algebra enlarges the space of mass matrices while leaving their attainable spectra unchanged. The projector contraction changes this result. At a particular fixed ratio of the two couplings, both successive mass ratios in a single sector can grow without bound along a real vacuum path. The fit to the observed masses addresses a separate question. It establishes that the two contractions can reproduce both spectra with a single real Higgs vacuum, using different coupling ratios in the two sectors.

Babu, Wu and Xu obtained fits to fermion masses and mixing using general $S_3$-invariant Yukawa couplings in a three-doublet model~\cite{BabuWuXu2024}. Here the Clifford construction fixes the family tensors of both contractions. The two parameters specifying the common vacuum direction, together with the two sector coupling ratios, allow the four mass ratios to vary independently near the certified solution. There is therefore no local relation among those ratios on this branch. The fixed tensors do, however, have a consequence for neutral flavour change. For any nonzero pair of couplings in a given sector, no common choice of left- and right-handed fermion bases diagonalises all three Yukawa matrices. At the fitted vacuum, the light Higgs couples in proportion to the fermion masses, so the off-diagonal couplings appear among the additional neutral scalars.

Requiring the fitted real direction to be stationary in the exact-$S_3$, CP-conserving tree-level potential~\cite{EmmanuelCostaEtAl2016} forces an accidental $SO(2)$ symmetry, leaving a flat direction. The quadratic soft term removes this flat direction by selecting the fitted vacuum and gives the additional doublets a mass that can be raised at fixed quartic coupling. For the reference fit and the degenerate heavy-scalar spectrum, kaon mixing gives a lower bound of about $318\,\mathrm{TeV}$ on the common mass, using leading-order QCD evolution and the adopted ten-percent allowance. The $2000\,\mathrm{TeV}$ benchmark satisfies the neutral-meson mixing and $\mu\to3e$ tests considered here. A dynamical account of the vacuum would still have to explain why the soft term selects the direction required by the mass fit.

Including the up-quark Yukawa sector would test whether the same vacuum can accommodate the up-quark masses and the relative orientation of the left-handed quark mass bases encoded in the CKM matrix. A radiative analysis would determine whether renormalisation generates Yukawa tensors outside the span of the two contractions and whether light-Higgs alignment and the neutral mass degeneracies responsible for the equal-chirality cancellation survive quantum corrections~\cite{FerreiraLavouraSilva2010,PenuelasPich2017,Chakrabarty2016,DarvishiMasouminiaPilaftsis2021}.

\section*{Statements and Declarations}

\paragraph{Funding}
No funds, grants, or other support were received.

\paragraph{Competing interests}
The authors have no relevant financial or non-financial interests to disclose.

\paragraph{Author contributions}
YZ led the analysis and prepared the initial draft, and NG developed the underlying Clifford-algebra model and contributed to the phenomenological calculations. Both authors helped shape the study and revised the manuscript.

\paragraph{Data availability}
The code and input files for the algebraic checks, both mass benchmarks, rational interval certificates, and scalar and neutral-flavour calculations are provided in the Supplementary Material. The mass inputs are specified in Sec.~\ref{Sec:Masses} and Appendix~\ref{App:Mass}.

\appendix

\section{Mass matrices and proofs}\label{App:Mass}

We specify the internal matrices and trace tensors before proving the spectral and existence statements of Sec.~\ref{Sec:Masses}. Matrix units $E_{ij}$ are numbered from 1 to 32. In the basis of Refs.~\cite{Gresnigt2026Technical,Gresnigt2026Higgs}, let $U=I_2\otimes U_{16}$ and $\mathcal E=I_2\otimes\operatorname{diag}(I_8,-I_8)$. On the ordered coordinate pairs $(1,14)$ and $(6,9)$, respectively, $U_{16}$ has the blocks $B_+$ and $B_-$; on each of $(2,10)$, $(3,11)$, $(4,12)$, $(5,13)$, $(7,15)$ and $(8,16)$ it has $B_0$, where
\begin{equation}
B_\pm=\frac12\begin{pmatrix}1&\pm i\sqrt3\\\pm i\sqrt3&1\end{pmatrix},\qquad B_0=\frac12\begin{pmatrix}1&\sqrt3\\-\sqrt3&1\end{pmatrix}.
\label{eq:mass-source-blocks}
\end{equation}
All other entries vanish. The family maps are $\psi(X)=UXU^\dagger$ and $\epsilon(X)=\mathcal EX\mathcal E$. Here $U^3=-I$, so $\psi^3=1$, while $\mathcal E U\mathcal E=U^\dagger$ gives $\epsilon\psi\epsilon=\psi^{-1}$. The charged-lepton and neutral-Higgs seeds are
\begin{equation}
L_1=-iE_{25,17},\qquad R_1=iE_{25,9},\qquad H_1=-E_{17,9}.
\label{eq:mass-source-seeds}
\end{equation}
For $a=2,3$, use $\eta=+1,-1$, respectively, to write their images as
\begin{align}
L_a&=-\frac i4E_{25,17}-\frac{\eta\sqrt3}{4}(E_{22,17}+E_{25,30})+\frac{3i}{4}E_{22,30},\nonumber\\
R_a&=\frac i4E_{25,9}+\frac{\eta\sqrt3}{4}(E_{22,9}-E_{25,6})+\frac{3i}{4}E_{22,6},\nonumber\\
H_a&=-\frac14E_{17,9}+\frac34E_{30,6}-\frac{\eta i\sqrt3}{4}(E_{17,6}+E_{30,9}).
\label{eq:mass-source-orbits}
\end{align}
These expressions give the Gram matrix $G=(3I+J)/4$ used in Sec.~\ref{Sec:Setup}. They also give
\begin{equation}
Y_1=\begin{pmatrix}1&1/4&1/4\\1/4&1/4&-1/8\\1/4&-1/8&1/4\end{pmatrix},\qquad Y_a=C_3^{a-1}Y_1(C_3^T)^{a-1},
\label{eq:mass-raw-old-tensors}
\end{equation}
where $C_3$ sends label 1 to 2, 2 to 3 and 3 to 1. The reflection, with $F$ exchanging labels 2 and 3, gives $\epsilon(L_a)=-L_{F(a)}$, $\epsilon(R_a)=R_{F(a)}$ and $\epsilon(H_a)=-H_{F(a)}$. Hence $\rho_L(\epsilon)Y_a\rho_R(\epsilon)^T=-Y_{F(a)}$, with the same transformation for $T_a$. The opposite left and right reflection signs in the down sector give the same tensor law. In particular, neither interaction replaces the sign-twisted Higgs orbit by an untwisted permutation triplet.

For an arbitrary $H\in\mathcal A$, only its entries $(17,9)$, $(17,6)$, $(30,9)$ and $(30,6)$ enter the trace: this follows directly from the support of $L_r^\dagger R_s$. Put $h_1=\overline H_{17,9}$, $h_2=\overline H_{17,6}$, $h_3=\overline H_{30,9}$ and $h_4=\overline H_{30,6}$. In the canonical frame of Sec.~\ref{Sec:Setup}, the resulting matrix is the one in Eq.~\eqref{eq:mass-right-image}, with
\begin{align}
d&=\frac{h_4-h_1}{2},&t&=\frac{i(h_3-h_2)}2,\nonumber\\
p&=-\frac{\sqrt3(h_1+h_4)}4+\frac{i(h_2+h_3)}4,\nonumber\\
q&=-\frac{h_1+h_4}4-\frac{i\sqrt3(h_2+h_3)}4.
\label{eq:mass-right-support-map}
\end{align}
This map is onto: its inverse is
\begin{align}
h_1&=-d-\frac{\sqrt3p+q}{2},&h_4&=d-\frac{\sqrt3p+q}{2},\nonumber\\
h_2&=-\frac{i(p-\sqrt3q)}2+it,&h_3&=-\frac{i(p-\sqrt3q)}2-it.
\label{eq:mass-right-support-inverse}
\end{align}
The transformation of a raw coefficient matrix is $Y\mapsto D_GQ^TYQD_G$, with $Q$ and $D_G$ fixed in Sec.~\ref{Sec:Setup}. Canonical normalisation of Eq.~\eqref{eq:mass-raw-old-tensors} gives $O(z)$ in Eq.~\eqref{eq:mass-two-matrices}. As a matrix space it is exactly the equal-diagonal arrow space $t=0$, although its diagonal coordinate is $2c$ when expressed in the common scalar coordinates of Eq.~\eqref{eq:mass-vacuum-coordinates}.

The fixed projector in Eq.~\eqref{eq:mass-projectors} is diagonal in this source basis. Its entries are one on the index set $\{1,5,7,8,9,10,11,12\}$ and on its translate by 16, and zero elsewhere. Thus $A_1^\dagger=A_1=A_1^2$ and $\operatorname{rank}A_1=16$; equivalently $Z_1^2=I_{32}/4$. The source expression for $Q_1$ commutes with the colour generators, which act among $a_1,a_2,a_3$, and with the weak and hypercharge generators~\cite{Gresnigt2026Technical,Gresnigt2026Higgs}. The family automorphisms fix these gauge generators, so all the $A_a$ commute with them. Conjugation by $U$ and $\mathcal E$ gives the stated family orbit of the projectors. Multiplying Eqs.~\eqref{eq:mass-source-seeds} and \eqref{eq:mass-source-orbits} by $A_1$ verifies $A_1L_rH_1=G_{1r}R_1$; applying $\psi$ proves Eq.~\eqref{eq:mass-new-tensor} for every $a$. Summing the outer products gives $\sum_a z_aT_a=G\operatorname{diag}(z)G$, whose normalisation yields $N(z)$ in Eq.~\eqref{eq:mass-two-matrices}. The inverse scalar map, useful below, is
\begin{equation}
z_{1,2}=\frac{4c+2q\pm2\sqrt3p}{3},\qquad z_3=\frac{4(c-q)}3.
\label{eq:mass-inverse-vacuum}
\end{equation}
For one colour the down seeds are $d_{L,1}=iE_{8,17}$ and $d_{R,1}=iE_{8,9}$. Their $U$ orbits have the same Gram matrix. The identical multiplication gives $-Y_a$ and $-T_a$; the other colours, with seed rows 5 and 7, give the same tensors, and cross-colour traces vanish. This establishes the common down-sector sign used in Sec.~\ref{Sec:Masses}.

For clarity, these are coefficient-field interactions, with ordinary four-dimensional antifermions and Hermitian conjugate terms. Let $\Pi_R$ denote the Hilbert--Schmidt projection onto $\operatorname{span}\{R_1,R_2,R_3\}$. The original right action need not map the selected left-handed states into this right-handed span. Indeed,
\begin{align}
L_2H_1&=\frac i4E_{25,9}+\frac{\sqrt3}{4}E_{22,9},\nonumber\\
(1-\Pi_R)(\alpha I+\beta A_1)L_2H_1&=\frac{\sqrt3\alpha}{8}(E_{22,9}+E_{25,6}).
\label{eq:mass-closure-distinction}
\end{align}
The residual is orthogonal to all $R_s$ and therefore does not enter the traced Yukawa tensor. The first identity in Eq.~\eqref{eq:mass-new-tensor} does not imply pointwise closure for a mixed vertex on all source-labelled states. The internal map corresponding to Eq.~\eqref{eq:mass-pencil} is $\sum_a\overline z_a[\overline\lambda_f XH_a+\overline\mu_f A_aXH_a]$. The dagger in the trace restores $\lambda_fz_a$ and $\mu_fz_a$ exactly once. This fixes both the scalar-dual and coupling conjugation conventions.

We next prove the spectral statement in Eq.~\eqref{eq:mass-right-bound}. Conjugating a matrix in Eq.~\eqref{eq:mass-right-image} by $\operatorname{diag}(i,1,1)$ gives $dI+A$ with $A^T=-A$. Write $A=[u+iv]_\times$, where $u,v\in\mathbb R^3$ and $[w]_\times x=w\times x$. Choose a real unit vector perpendicular to both $u$ and $v$, and take it as the first column of a proper real orthogonal matrix. In this frame the 23 entry of $A$ vanishes. A further diagonal unitary similarity makes its remaining two antisymmetric pairs symmetric. Thus every matrix in the full image has the singular spectrum of
\begin{equation}
B=\begin{pmatrix}d&p&q\\p&d&0\\q&0&d\end{pmatrix}.
\label{eq:mass-arrow-proof}
\end{equation}
Conversely, every such $B$ belongs to the image. These matrix-dependent transformations establish a one-sector spectral equivalence; they are not a common flavour-basis change for two sectors.

First suppose the ordered masses $a\le b\le h$ are positive. Define $S=a^2+b^2+h^2$, $E=a^2b^2+a^2h^2+b^2h^2$ and $P=abh$. For Eq.~\eqref{eq:mass-arrow-proof}, set $C=|d|^2$, $X=C+|p|^2+|q|^2$ and $\zeta=d^2-p^2-q^2$. The trace, determinant and second characteristic coefficient of $BB^\dagger$ give
\begin{equation}
S=C+2X,\qquad P^2=C|\zeta|^2,\qquad E=X^2+2\operatorname{Re}(d^2\overline\zeta).
\label{eq:mass-arrow-invariants}
\end{equation}
The last identity follows by summing the squared moduli of the $2\times2$ minors. The lower principal block of $BB^\dagger$ has eigenvalues $C$ and $X$, so interlacing gives $a^2\le C\le b^2\le X\le h^2$~\cite{HornJohnson2012}. Equation~\eqref{eq:mass-arrow-invariants} then implies $(X^2-E)^2\le4P^2(S-2X)$. The difference factors as
\begin{align}
&(X^2-E)^2-4P^2(S-2X)\nonumber\\
&\hspace{1em}=(X-r_0)(X-r_1)(X-X_-)(X-X_+),\nonumber\\
r_0&=-ab-ah-bh,\qquad r_1=ab+ah-bh,\nonumber\\
X_\pm&=bh\pm a(h-b).
\label{eq:mass-cone-factorization}
\end{align}
Here $r_0<0$ and $r_1-a^2=-(b-a)(h-a)\le0$. Interlacing therefore excludes the lower sign interval of this quartic, leaving $X_-\le X\le X_+$. If $X=r_1$ is a possible repeated-root endpoint, interlacing forces $a=b$ and it already equals $X_-$. Using $C=S-2X$, the necessary condition is
\begin{equation}
[a^2,b^2]\cap[(h-a-b)^2,(h-b+a)^2]\ne\varnothing.
\label{eq:mass-cone-interval}
\end{equation}
The second upper endpoint is at least $a^2$, and its lower endpoint is no larger than its upper endpoint. Thus the only remaining condition is $|h-a-b|\le b$, or $h\le a+2b$.

For sufficiency choose $C$ in the intersection in Eq.~\eqref{eq:mass-cone-interval}, put $d=\sqrt C$, $X=(S-C)/2$, $R=X-C$, and choose
\begin{equation}
\zeta=\frac{P}{\sqrt C}e^{i\delta},\qquad \cos\delta=\frac{E-X^2}{2P\sqrt C}.
\label{eq:mass-cone-phase}
\end{equation}
Equation~\eqref{eq:mass-cone-factorization} ensures that the cosine lies in $[-1,1]$. Writing $w=C-\zeta=\rho e^{i\varphi}$ gives
\begin{equation}
R^2-|w|^2=\frac{(C-a^2)(C-b^2)(C-h^2)}{C}\ge0.
\label{eq:mass-cone-compatibility}
\end{equation}
Also $X\ge X_-=b^2+(b-a)(h-b)\ge b^2\ge C$, so $R\ge0$. Consequently the choice
\begin{equation}
p=e^{i\varphi/2}\sqrt{\frac{R+\rho}{2}},\qquad q=i e^{i\varphi/2}\sqrt{\frac{R-\rho}{2}}
\label{eq:mass-cone-construction}
\end{equation}
has $p^2+q^2=w$ and $|p|^2+|q|^2=R$; take $\varphi=0$ if $w=0$. It reproduces all three invariants in Eq.~\eqref{eq:mass-arrow-invariants}, and hence exactly the required squared singular values. The arrow matrix is complex symmetric, so these singular values can equivalently be viewed as its nonnegative Takagi values~\cite{Youla1961}; the feasibility condition above is derived directly, rather than assumed from a general diagonal-entry theorem. This argument includes positive repeated masses. For zero target masses, replace each mass by $m_i+\varepsilon$: ordering and the cone inequality are preserved, and the constructed matrices have bounded Frobenius norm. A convergent subsequence as $\varepsilon\to0$ realises the limiting spectrum. Conversely, a given singular arrow can be perturbed by arbitrarily small scalar multiples of $I$ to make its determinant nonzero; the determinant is a monic cubic in that perturbation. Necessity then follows by continuity. This proves the entire closed cone, including its boundary.

We now classify the fixed-coupling families in Eq.~\eqref{eq:mass-pencil}. Put $g=\lambda+\mu$ and $b=2\lambda+\mu$. In the common scalar coordinates,
\begin{equation}
M=\begin{pmatrix}2gc&gp&gq\\gp&bc+\mu q/2&\mu p/2\\gq&\mu p/2&bc-\mu q/2\end{pmatrix}.
\label{eq:mass-coexistence-matrix}
\end{equation}
Its three coefficient matrices are Frobenius-orthogonal, giving
\begin{align}
\lVert M\rVert_F^2={}&(4|g|^2+2|b|^2)|c|^2\nonumber\\
&+2\left(|g|^2+\frac{|\mu|^2}{4}\right)(|p|^2+|q|^2).
\label{eq:mass-coexistence-norm}
\end{align}
Both coefficients are positive for every nonzero coupling pair, so the map from $(c,p,q)$ to $M$ is injective.

Suppose first that $gb\ne0$. A nonzero complex symmetric rank-one matrix can be written as $vv^T$. If $c=0$, its 11 entry forces $v_1=0$ and hence $gp=gq=0$, making $M=0$. We may therefore set $c=1$. The Schur complement of the 11 entry vanishes precisely when
\begin{equation}
gp^2=2b+\mu q,\qquad gpq=\mu p,\qquad gq^2=2b-\mu q.
\label{eq:mass-rank-one-schur}
\end{equation}
For $p\ne0$, the middle equation gives $q=\mu/g$; the other two then give $p^2=3b/g$ and $\mu^2=bg$. For $p=0$, the first equation forces $\mu\ne0$ and gives $q=-2b/\mu$, and the last again gives $\mu^2=bg$. Thus rank one occurs in this chart exactly when
\begin{equation}
\mu^2=bg\quad\Longleftrightarrow\quad\lambda(2\lambda+3\mu)=0.
\label{eq:mass-rank-one-condition}
\end{equation}
Conversely, this condition gives the rank-one points $(p,q)=(\pm\sqrt{3b/g},\mu/g)$ and $(0,-2\mu/g)$ at $c=1$.

The remaining charts must be treated separately. If $b=0$, then $\lambda/\mu=-1/2$; setting $\mu=1$, two $2\times2$ minors of Eq.~\eqref{eq:mass-coexistence-matrix} are $-(p^2+q^2)/4$ and $(p^2-q^2)/4$. Rank one forces $p=q=0$, giving $\operatorname{diag}(c,0,0)$ for $c\ne0$. If $g=0$, then $\lambda/\mu=-1$: the first row and column vanish, and the lower determinant vanishes exactly when $p^2+q^2=4c^2$. Every nonzero point on this complex conic has rank one, but no member of the family has rank three. The pure original interaction $\mu=0$ is covered by the first chart and fails Eq.~\eqref{eq:mass-rank-one-condition}. This proves Eq.~\eqref{eq:mass-rank-one-ratios} for all nonzero complex coupling pairs.

The determinant is
\begin{align}
\det M=g\biggl[&2b^2c^3-\left(\frac{\mu^2}{2}+bg\right)c(p^2+q^2)\nonumber\\
&+\frac{g\mu}{2}(3p^2q-q^3)\biggr].
\label{eq:mass-coexistence-determinant}
\end{align}
For $gb\ne0$, $(c,p,q)=(1,0,0)$ gives full rank. For $g\ne0$, $b=0$, the point $(0,0,1)$ has determinant $-\mu^3/8\ne0$. Thus only $g=0$ makes the entire family singular. At fixed couplings, normalise $|c|^2+|p|^2+|q|^2=1$. Equation~\eqref{eq:mass-coexistence-norm} ensures a compact nonzero image. If it contains no rank-one matrix, its second singular value has a positive minimum, and $m_3/m_2$ is bounded. Conversely, an unbounded-ratio sequence has a nonzero rank-one limit. Whenever both a rank-one member $M_1$ and a full-rank member $M_3$ exist, $M_1+\varepsilon M_3$ is full rank for arbitrarily small $\varepsilon$: its determinant is a nonzero polynomial with leading coefficient $\det M_3$. These matrices approach rank one and have unbounded $m_3/m_2$. Among full-rank matrices, $m_3/m_2$ is therefore unbounded at fixed couplings precisely when $\mu\ne0$ and $\lambda/\mu\in\{0,-1/2,-3/2\}$.

For the real path in Eq.~\eqref{eq:mass-hierarchy-path}, put $w=(\varepsilon^2,\varepsilon,1)^T$ and
\begin{equation}
V=\begin{pmatrix}1&1&1\\-\sqrt3&\sqrt3&0\\-1&-1&2\end{pmatrix}.
\label{eq:mass-path-factor}
\end{equation}
Direct substitution at $(\lambda,\mu)=(-3/2,1)$ gives $\widetilde M=M/v_*=-V\operatorname{diag}(w)V^T$ and $V^TV=6I-J$. Thus $-\widetilde M$ is positive definite, with characteristic invariants
\begin{equation}
\begin{aligned}
\operatorname{tr}(-\widetilde M)&=5\sum_iw_i,\\
e_2(-\widetilde M)&=24\sum_{i<j}w_iw_j,\qquad\det(-\widetilde M)=108\prod_iw_i.
\end{aligned}
\label{eq:mass-path-invariants}
\end{equation}
Here $e_2(B)$ denotes the sum of its principal $2\times2$ minors. In units of $v_*$, the heavy mass tends to 5. The determinant then gives $(m_1/v_*)(m_2/v_*)\sim108\varepsilon^3/5$, while $e_2$ gives $(m_1+m_2)/v_*\sim24\varepsilon/5$. Their product divided by the squared sum tends to zero, so $m_1/m_2\to0$. It follows that $m_2/v_*\sim24\varepsilon/5$ and $m_1/v_*\sim9\varepsilon^2/2$, proving Eq.~\eqref{eq:mass-hierarchy-asymptotics}.

To prove Eq.~\eqref{eq:mass-proportionality}, suppose $O(c,p,q)=N(C,P,Q')$. The 23 and 12 entries give $P=p=0$, the 13 entry gives $q=Q'$, and the 11 entry gives $c=C$. Comparing the last two diagonal entries then forces $Q'=c=0$. Hence the two three-dimensional tensor spaces have zero intersection. For each fixed nonzero $z$, the matrices $O(z)$ and $N(z)$ are nonzero and linearly independent, which proves the equivalence. This statement concerns matrices, not their spectra: at $c=1$, $p=q=0$, the coupling pairs $(1,1)$ and $(-5,7)$ give $\operatorname{diag}(4,3,3)$ and $\operatorname{diag}(4,-3,-3)$, with equal singular spectra but nonproportional matrices.

We finish with the exact common-vacuum construction. In the real chart $c(\widehat z)=1$, the dimensionless matrix is
\begin{equation}
F_r=\begin{pmatrix}2(r+1)&(r+1)p&(r+1)q\\(r+1)p&2r+1+q/2&p/2\\(r+1)q&p/2&2r+1-q/2\end{pmatrix}.
\label{eq:mass-joint-matrix}
\end{equation}
Put $R=p^2+q^2$, $T=q(q^2-3p^2)$, $a=2(r+1)$, $b=2r+1$, $g=r+1$ and $h=6r+4$. Its characteristic coefficients are
\begin{align}
\operatorname{tr}F_r&=h,\nonumber\\
e_2(F_r)&=b^2+2ab-(g^2+1/4)R,\nonumber\\
\det F_r&=ab^2-(a/4+bg^2)R-g^2T/2.
\label{eq:mass-joint-invariants}
\end{align}
The real domain is exactly $R\ge0$, $T^2\le R^3$. For $R>0$, write $q=\sqrt R\cos\alpha$, $p=\sqrt R\sin\alpha$; then $T=R^{3/2}\cos3\alpha$, which proves both directions. The endpoint $R=0$ has $p=q=T=0$. In particular, $R^3-T^2=p^2(p^2-3q^2)^2$.

The four exact target ratios from Eq.~\eqref{eq:mass-inputs} are
\begin{align}
x_e&=\frac{243285077}{873085000000},&y_e&=\frac{1027181337}{17461700000},\nonumber\\
x_d&=\frac{141}{143000},&y_d&=\frac{57}{2860}.
\label{eq:mass-exact-targets}
\end{align}
Choose signed ratios $(\xi_e,\eta_e)=(x_e,y_e)$ and $(\xi_d,\eta_d)=(-x_d,y_d)$, and define $u_{1f}=1+\xi_f+\eta_f$, $u_{2f}=\xi_f+\eta_f+\xi_f\eta_f$ and $u_{3f}=\xi_f\eta_f$. A negative light down eigenvalue is allowed because physical masses are singular values. With $\theta=(p,q,r_e,r_d)$ and $\rho=(x_e,y_e,x_d,y_d)$, define the four-component rational polynomial $\mathcal P=(\mathcal P_{2e},\mathcal P_{3e},\mathcal P_{2d},\mathcal P_{3d})^T$ by
\begin{align}
\mathcal P_{2f}&=u_{1f}^2e_2(F_{r_f})-h_f^2u_{2f},\nonumber\\
\mathcal P_{3f}&=u_{1f}^3\det F_{r_f}-h_f^3u_{3f},\qquad h_f=6r_f+4.
\label{eq:mass-joint-polynomials}
\end{align}
At a zero with $h_f/u_{1f}>0$, the characteristic polynomial has exactly the signed eigenvalues $h_f(\xi_f,\eta_f,1)/u_{1f}$. The positive sector factors
\begin{equation}
\widehat\mu_e=\frac{m_\tau u_{1e}}{h_e},\qquad\widehat\mu_d=\frac{m_bu_{1d}}{h_d}
\label{eq:mass-exact-scales}
\end{equation}
then reproduce all six benchmark masses. Thus it suffices to prove a zero of four explicit polynomials, without rounding the target ratios or diagonalising an approximate matrix.

A rational certificate specifies such a zero. Every terminating decimal in the certificate is treated as an exact rational number. Let $\rho_0$ be Eq.~\eqref{eq:mass-exact-targets}, and set
\begin{equation}
\begin{gathered}
m=\begin{pmatrix}0.022393015646750963516254317477\\-1.995680018694952986833502266600\\0.130485747369318996810678287340\\0.036446861328347846415094950641\end{pmatrix},\\
C=[D_\theta\mathcal P(m;\rho_0)]^{-1}.
\end{gathered}
\label{eq:mass-certificate-centre}
\end{equation}
The inverse exists by exact rational elimination. For $X=m+[-10^{-24},10^{-24}]^4$, exact interval evaluation gives
\begin{align}
\lVert C\mathcal P(m;\rho_0)\rVert_\infty&<5\times10^{-31},\nonumber\\
\sup_{\theta\in X}\lVert I-CD_\theta\mathcal P(\theta;\rho_0)\rVert_\infty&<2\times10^{-22}.
\label{eq:mass-point-certificate}
\end{align}
These bounds are obtained by substituting $\theta=m+\delta\theta$ into the full polynomials, combining identical monomials, and enclosing their values with rational interval arithmetic. No Taylor terms are discarded. The polynomials, centre, box and exact inverse above completely specify the calculation; the certification uses the inclusion principles of interval analysis and verified numerical computation~\cite{MooreKearfottCloud2009,Rump2010}. The Supplementary Material reconstructs the source tensors and reproduces these rational certificates with the exact centre in Eq.~\eqref{eq:mass-certificate-centre}, including an independently implemented certificate checker.

Define $\mathcal T_\rho(\theta)=\theta-C\mathcal P(\theta;\rho)$. At $\rho=\rho_0$, the segment integral formula and Eq.~\eqref{eq:mass-point-certificate} imply, for $\theta\in X$,
\begin{equation}
\lVert\mathcal T_{\rho_0}(\theta)-m\rVert_\infty<5\times10^{-31}+2\times10^{-46}<10^{-24}.
\label{eq:mass-point-inclusion}
\end{equation}
Thus $\mathcal T_{\rho_0}$ maps $X$ into its interior and is a contraction there. Banach's fixed-point theorem gives a unique fixed point in $X$. Since $C$ is invertible, it is exactly a zero of $\mathcal P$. Throughout this box, $h_f/u_{1f}>0$ and $0<x_f<y_f<1$, so Eq.~\eqref{eq:mass-exact-scales} gives the required ordered nonzero masses. The exact solution is this certified root; the values printed in Sec.~\ref{Sec:Masses} are its rounded representatives.

The same argument gives a finite neighbourhood of realisable targets. This is a parameter-dependent enclosure problem of the type studied in rigorous sensitivity analysis~\cite{Neumaier1989}; here all bounds are computed with exact rational intervals. Let $\mathcal U$ be the box in which the four positive ratios vary independently about $\rho_0$ with relative half-widths $(10^{-8},10^{-8},0.01,0.01)$. Set
\begin{equation}
\begin{gathered}
H=(0.0005,0.00015,0.00012,0.001)^T,\\
X_H=m+\prod_{i=1}^4[-H_i,H_i],\qquad D_H=\operatorname{diag}(H).
\end{gathered}
\label{eq:mass-uniform-box}
\end{equation}
Write $Z_\theta=I-CD_\theta\mathcal P(\theta;\rho)$, $v_i=\sup_{\rho\in\mathcal U}|(C\mathcal P(m;\rho))_i|$ and $B_{ij}=\sup_{X_H\times\mathcal U}|(Z_\theta)_{ij}|$. Full polynomial translation in both parameters and targets gives the exact bounds
\begin{align}
\sup_{X_H\times\mathcal U}\lVert D_H^{-1}Z_\theta D_H\rVert_\infty&<0.062,\nonumber\\
\max_i\frac{v_i+\sum_j B_{ij}H_j}{H_i}&<0.565.
\label{eq:mass-uniform-certificate}
\end{align}
The second inequality places $\mathcal T_\rho(X_H)$ strictly inside $X_H$, and the first makes it a contraction in the weighted infinity norm. Hence every target $\rho\in\mathcal U$ gives a unique zero of $\mathcal P(\cdot;\rho)$ in $X_H$. The same rational bounds give $R^3-T^2>0.06839$ throughout $X_H$, and both $r_f$ remain positive. Target ordering and $h_f/u_{1f}>0$ also hold uniformly. The common vacuum and couplings vary with the target, while the two heavy masses are held fixed.

Finally, the contraction estimate makes $D_\theta\mathcal P$ invertible at the solution, since $CD_\theta\mathcal P$ differs from $I$ by a matrix of norm less than one. For each sector,
\begin{equation}
\det\frac{\partial(u_2/u_1^2,u_3/u_1^3)}{\partial(\xi,\eta)}=\frac{(\xi-1)(\xi-\eta)(\eta-1)}{(1+\xi+\eta)^6}\ne0.
\label{eq:mass-ratio-jacobian}
\end{equation}
To see how this enters the four-equation system, divide $\mathcal P_{2f}$ and $\mathcal P_{3f}$ by $u_{1f}^2h_f^2$ and $u_{1f}^3h_f^3$, respectively. The resulting equations are $e_2(F_{r_f})/h_f^2-u_{2f}/u_{1f}^2=0$ and $\det F_{r_f}/h_f^3-u_{3f}/u_{1f}^3=0$. At a zero, this row rescaling preserves derivative ranks. Equation~\eqref{eq:mass-ratio-jacobian}, together with the fixed sign choices, makes $D_\rho\mathcal P$ invertible. Implicit differentiation gives $D_\theta\rho=-(D_\rho\mathcal P)^{-1}D_\theta\mathcal P$, which is nonsingular. The four physical mass ratios therefore vary independently in a neighbourhood of the certified common-vacuum solution.

For the updated comparison in Eq.~\eqref{eq:mass-updated-benchmark}, the quoted Standard Model parameters at $M_Z$ are~\cite{AntuschHinzeSaad2026}
\begin{equation}
\begin{gathered}
(y_e,y_\mu,y_\tau)=10^{-6}(2.77713,585.042,9937.8),\\
(y_d,y_s,y_b)=10^{-3}(0.0154,0.306,16.30),\\
\bar v(M_Z)=248.401\,\mathrm{GeV}.
\end{gathered}
\label{eq:mass-updated-inputs}
\end{equation}
The definition $m_f^{\mathrm{cmp}}=y_f\bar v/\sqrt2$ fixes the two heavy mass scales, while the common $\bar v$ cancels from the four exact ratio targets,
\begin{equation}
\rho_{\mathrm u}=\left(\frac{30857}{110420000},\frac{97507}{1656300},\frac{77}{81500},\frac{153}{8150}\right).
\label{eq:mass-updated-targets}
\end{equation}
Use the same polynomial system and signed branches as in Eq.~\eqref{eq:mass-joint-polynomials}, with rational centre
\begin{equation}
m_{\mathrm u}=\begin{pmatrix}
0.021095238023830260739952726170\\
-1.996007106575430565513074956060\\
0.130873144240804143351777602830\\
0.034277029276454023613244231599
\end{pmatrix}.
\label{eq:mass-updated-centre}
\end{equation}
Writing $C_{\mathrm u}=[D_\theta\mathcal P(m_{\mathrm u};\rho_{\mathrm u})]^{-1}$ and $X_{\mathrm u}=m_{\mathrm u}+[-10^{-24},10^{-24}]^4$, exact rational evaluation gives
\begin{equation}
\begin{gathered}
\|C_{\mathrm u}\mathcal P(m_{\mathrm u};\rho_{\mathrm u})\|_\infty<4\times10^{-31},\\
\sup_{\theta\in X_{\mathrm u}}\|I-C_{\mathrm u}D_\theta\mathcal P(\theta;\rho_{\mathrm u})\|_\infty<2\times10^{-22}.
\end{gathered}
\label{eq:mass-updated-certificate}
\end{equation}
For the same shape half-widths $H$ as in Eq.~\eqref{eq:mass-uniform-box} and relative target half-widths $(10^{-8},10^{-8},0.01,0.01)$ about $\rho_{\mathrm u}$, the weighted derivative and self-mapping bounds defined in Eq.~\eqref{eq:mass-uniform-certificate} are less than $0.064$ and $0.528$, respectively. The sign, ordering and nonexceptional-angle conditions remain strict throughout the boxes. The contraction argument therefore supplies a unique physical solution in the stated parameter box for each target in this neighbourhood, and the argument following Eq.~\eqref{eq:mass-ratio-jacobian} again establishes local independence of the four mass ratios. These calculations certify the stated fixed targets and feasibility neighbourhood; the extra digits do not imply additional precision in the source inputs. The Supplementary Material reproduces both certificates and the associated scalar and flavour results.

\section{Scalar and flavour calculations}\label{App:Scalar}

For completeness, we give an explicit matrix convention for the internal Higgs construction in Sec.~\ref{Sec:Setup}. Let $\tau_0=I_2$, let $\tau_{1,2,3}$ be the Pauli matrices, and write $\tau_{ijkl}=\tau_i\otimes\tau_j\otimes\tau_k\otimes\tau_l$. Define
\begin{equation}
\begin{gathered}
\begin{aligned}
e_1&=i\tau_{1111},&e_2&=i\tau_{1130},\\
e_3&=-i\tau_{1113},&e_4&=-i\tau_{1300},\\
e_5&=-i\tau_{1112},&e_6&=-i\tau_{1200},\\
e_7&=i\tau_{1120},&e_8&=i\tau_{2000},
\end{aligned}\\
\omega=e_1\cdots e_8=\operatorname{diag}(I_8,-I_8).
\end{gathered}
\label{eq:setup-Clifford-matrices}
\end{equation}
These matrices obey $e_i^\dagger=-e_i$ and $\{e_i,e_j\}=-2\delta_{ij}I_{16}$. The five Witt generators in the $32$-dimensional representation are
\begin{equation}
\begin{gathered}
a_j=I_2\otimes\frac{-e_j+ie_{j+4}}2\quad(j=1,\ldots,4),\\
a_5=\frac{i\tau_1-\tau_2}{2}\otimes\omega.
\end{gathered}
\label{eq:setup-Witt-matrices}
\end{equation}
They give the seeds in Eqs.~\eqref{eq:setup-lepton-seeds} and \eqref{eq:setup-Higgs-seeds}; the family lift is specified in Appendix~\ref{App:Mass}. In particular, $H_d^0=-E_{17,9}$ in its matrix-unit convention. This fixes all component phases without requiring the reader to reconstruct the original Clifford basis from an external source.

The weak generators can be checked directly. With $\mathcal E=I_2\otimes\omega$ and
\begin{equation}
P_W=f_{+++++}+f_{---++}+f_{++++-}+f_{---+-},
\label{eq:setup-weak-projector}
\end{equation}
they are
\begin{equation}
\begin{aligned}
T_1&=\frac{i(a_5^\dagger-a_5)\mathcal EP_W}{2},\\
T_2&=\frac{(a_5+a_5^\dagger)\mathcal EP_W}{2},\\
T_3&=\frac{(a_5^\dagger a_5-a_5a_5^\dagger)P_W}{2}.
\end{aligned}
\label{eq:setup-weak-generators}
\end{equation}
The electromagnetic and hypercharge generators are
\begin{equation}
Q_{\rm em}=\overline Q+(2P_W-I_{32})a_5^\dagger a_5,\qquad Y=2(Q_{\rm em}-T_3),
\label{eq:setup-electroweak-charges}
\end{equation}
where $\overline Q$ is the cyclic charge average defined in Sec.~\ref{Sec:Masses}. In particular, this average alone is not the full electromagnetic generator. The colour algebra is spanned by the traceless Hermitian combinations of $a_i^\dagger a_j$, with $i,j\in\{1,2,3\}$. These definitions reproduce the gauge weights of the twelve Higgs components and the gauge-compatible projectors used in the mass analysis~\cite{Gresnigt2026Technical,Gresnigt2026Higgs}.

In the original component convention, $[T_+,H_t^{\rm lower}]=-iH_t^{\rm upper}$ and $[T_-,H_t^{\rm upper}]=iH_t^{\rm lower}$, with $T_\pm=T_1\pm iT_2$. The phases in Eq.~\eqref{eq:setup-internal-doublets} convert both pairs to the standard weak-doublet basis. On the full twelve-dimensional Higgs space define
\begin{equation}
\mathcal C(X)=(a_5^\dagger+a_5)X^*(a_5^\dagger-a_5).
\label{eq:setup-antilinear-map}
\end{equation}
Here $X^*$ is entrywise complex conjugation in the matrix convention above. Direct multiplication gives, on every family orbit,
\begin{equation}
\begin{aligned}
\mathcal C(H_d^0)&=H_u^0,&\mathcal C(H_d^-)&=-H_u^+,\\
\mathcal C(H_u^0)&=-H_d^0,&\mathcal C(H_u^+)&=H_d^-.
\end{aligned}
\label{eq:setup-antilinear-components}
\end{equation}
Thus $\mathcal C^2=-1$ on this space. In the standard bases of Eq.~\eqref{eq:setup-internal-doublets}, it sends the first basis vector to minus the second of the other doublet, and the second to the first. Consequently its action on either coefficient column is $\varphi\mapsto i\sigma_2\varphi^*$. The choice $\varphi_{d,a}=\widetilde\Phi_a$, $\varphi_{u,a}=\Phi_a$ then agrees with this identification up to the constant sign from $\widetilde{\widetilde\Phi}_a=-\Phi_a$; that sign is included in the sector-coupling convention.

The same matrices give $\mathcal C\psi=\psi\mathcal C$, $\mathcal C\epsilon=\epsilon\mathcal C$ and
\begin{equation}
\mathcal C([T,X])=-[T,\mathcal C(X)]\qquad\text{for }T=T_1,T_2,T_3,Y
\label{eq:setup-antilinear-gauge}
\end{equation}
on all Higgs components. Because $\mathcal C$ is antilinear, this last identity intertwines the finite gauge actions. It establishes compatibility of the chosen coefficient pairing with gauge and family symmetry. The map is neither Hermitian conjugation nor an algebra automorphism, and its use imposes no spacetime charge-conjugation symmetry or unique scalar kinetic prescription.

We write the exact-$S_3$ scalar potential in the family frame of Eq.~\eqref{eq:scalar-family-frame}. Define
\begin{equation}
\begin{aligned}
r_D&=D_1^\dagger D_1+D_2^\dagger D_2,\\
A_D&=D_1^\dagger D_1-D_2^\dagger D_2,\\
B_D&=D_1^\dagger D_2+D_2^\dagger D_1.
\end{aligned}
\label{eq:scalar-bilinears}
\end{equation}
With real coefficients and standard CP, the general renormalisable potential is, up to an additive constant~\cite{DasDey2014,EmmanuelCostaEtAl2016},
\begin{align}
V_{S_3}={}&m_S^2S^\dagger S+m_D^2r_D+\kappa_1r_D^2\nonumber\\
&+\kappa_2(D_1^\dagger D_2-D_2^\dagger D_1)^2\nonumber\\
&+\kappa_3(A_D^2+B_D^2)\nonumber\\
&+\kappa_4\bigl[(S^\dagger D_1)B_D+(S^\dagger D_2)A_D+\mathrm{h.c.}\bigr]\nonumber\\
&+\kappa_5(S^\dagger S)r_D\nonumber\\
&+\kappa_6\bigl[(S^\dagger D_1)(D_1^\dagger S)+(S^\dagger D_2)(D_2^\dagger S)\bigr]\nonumber\\
&+\kappa_7\bigl[(S^\dagger D_1)^2+(S^\dagger D_2)^2+\mathrm{h.c.}\bigr]\nonumber\\
&+\kappa_8(S^\dagger S)^2.
\label{eq:scalar-exact-potential}
\end{align}
Using the pseudosinglet convention of Ref.~\cite{EmmanuelCostaEtAl2016}, specifically Eqs.~(2.13a) and (2.13b) in Sec.~2.4, we identify $h_A=S$, $h_1=D_2$ and $h_2=-D_1$. With this identification, $\mu_0^2=m_S^2$, $\mu_1^2=m_D^2$ and $\lambda_i=\kappa_i$ for $i=1,\ldots,8$. In particular, no sign or normalisation change of $\kappa_4$ is implicit. The cited potential and stationary conditions are understood with their published errata. The eight quartic structures can also be obtained directly by averaging the 27 real CP-even quadratic monomials in the nine Hermitian family bilinears over $S_3$; the invariant subspace has dimension eight. The common sign in the twisted permutation representation cancels on these bilinears.

For the real neutral configuration in Eq.~\eqref{eq:scalar-family-frame}, put $R_D=x^2+y^2$, $\kappa_D=\kappa_1+\kappa_3$ and $K=\kappa_5+\kappa_6+2\kappa_7$. The potential reduces to
\begin{equation}
\begin{split}
V_0={}&\frac{m_S^2s^2+m_D^2R_D}{2}+\frac{\kappa_D}{4}R_D^2+\frac{K}{4}s^2R_D\\
&+\frac{\kappa_8}{4}s^4+\frac{\kappa_4}{2}sy(3x^2-y^2).
\end{split}
\label{eq:scalar-neutral-potential}
\end{equation}
Its derivatives are
\begin{align}
\partial_sV_0&=m_S^2s+\kappa_8s^3+\frac K2sR_D+\frac{\kappa_4}{2}y(3x^2-y^2),\nonumber\\
\partial_xV_0&=x\left[m_D^2+\kappa_DR_D+\frac K2s^2+3\kappa_4sy\right],\nonumber\\
\partial_yV_0&=y\left[m_D^2+\kappa_DR_D+\frac K2s^2\right]+\frac{3\kappa_4}{2}s(x^2-y^2).
\label{eq:scalar-stationarity}
\end{align}
Taking $y\partial_xV_0-x\partial_yV_0$ gives Eq.~\eqref{eq:scalar-angular-obstruction}. For the dimensionless mass coordinates of Sec.~\ref{Sec:Masses},
\begin{equation}
(s,x,y)=\frac{4v_*}{\sqrt3}(\sqrt2c,p,q).
\label{eq:scalar-vacuum-coordinates}
\end{equation}
The point certificate in Appendix~\ref{App:Mass} has $c=1$, $p\ne0$ and $p^2\ne3q^2$ throughout its exact rational box. Consequently $sx(3y^2-x^2)\ne0$, so stationarity forces $\kappa_4=0$ without relying on the rounded vacuum values.

At $\kappa_4=0$, every term of Eq.~\eqref{eq:scalar-exact-potential} is invariant under real rotations of $(D_1,D_2)$. Since $s\ne0$ and $R_D>0$, the remaining stationary equations are
\begin{equation}
m_S^2=-\kappa_8s^2-\frac K2R_D,\qquad m_D^2=-\kappa_DR_D-\frac K2s^2.
\label{eq:scalar-radial-stationarity}
\end{equation}
Let $\mathcal K$ be the Hessian of $V_0$ with respect to $(s,x,y)$. Substituting Eq.~\eqref{eq:scalar-radial-stationarity} gives
\begin{equation}
\mathcal K\begin{pmatrix}0\\y\\-x\end{pmatrix}=0.
\label{eq:scalar-angular-zero-mode}
\end{equation}
Equivalently, this follows by differentiating the continuous orbit of stationary points. This real angular fluctuation is orthogonal to the vacuum and changes the relative family expectation values; it is not the neutral gauge Goldstone, which lies in the imaginary direction parallel to the vacuum. Thus even a positive semidefinite Hessian would leave a physical flat direction. The fixed Yukawa tensors need not respect this accidental scalar $SO(2)$, so the argument is restricted to the stated tree-level potential.

For the soft completion in Eq.~\eqref{eq:scalar-soft-potential}, the radial part corresponds to $\kappa_1=\kappa_8=\lambda_V$, $\kappa_5=2\lambda_V$ and all other $\kappa_i=0$, with the symmetric quadratic terms $m_S^2=m_D^2=-\lambda_Vv^2$ before adding $\mathsf S$. To obtain the full quadratic spectrum, expand
\begin{equation}
\Phi_a^0=\frac{vn_a+\rho_a+i\eta_a}{\sqrt2},\qquad \Phi_a^+=\frac{\xi_a+i\zeta_a}{\sqrt2},
\label{eq:scalar-real-fields}
\end{equation}
where $\rho$, $\eta$, $\xi$ and $\zeta$ are canonically normalised real family vectors. Then
\begin{equation}
\sum_a\Phi_a^\dagger\Phi_a-\frac{v^2}{2}=v\,n^T\rho+\frac{\rho^T\rho+\eta^T\eta+\xi^T\xi+\zeta^T\zeta}{2}.
\label{eq:scalar-radial-expansion}
\end{equation}
Using $\mathsf Sn=0$, the quadratic terms are
\begin{equation}
V^{(2)}=\lambda_Vv^2(n^T\rho)^2+\frac12\left(\rho^T\mathsf S\rho+\eta^T\mathsf S\eta+\xi^T\mathsf S\xi+\zeta^T\mathsf S\zeta\right),
\label{eq:scalar-quadratic-expansion}
\end{equation}
which yields Eqs.~\eqref{eq:scalar-hessians} and \eqref{eq:scalar-spectrum}. There is one zero direction in each of $\eta$, $\xi$ and $\zeta$, and no zero direction in $\rho$. The $U(2)$ symmetry of this scalar potential acts on doublets with zero VEVs and produces no additional Goldstone modes.

For reproducibility, an orthonormal completion suitable for the fitted $n$ is
\begin{equation}
\begin{gathered}
u=\frac{\boldsymbol e_1-n_1n}{\sqrt{1-n_1^2}},\qquad
w=\frac{(0,n_3,-n_2)^T}{\sqrt{1-n_1^2}},\\
R=\begin{pmatrix}n^T\\u^T\\w^T\end{pmatrix},\qquad \boldsymbol e_1=(1,0,0)^T.
\end{gathered}
\label{eq:scalar-higgs-rotation}
\end{equation}
Here $RR^T=I$ and $Rn=(1,0,0)^T$. The exact fitted vector defines $I-nn^T$; its rounded decimal entries should not be used to replace the exact kernel condition.

We next establish the common-diagonalisation obstruction used in Sec.~\ref{Sec:Flavour}. If independent unitaries make a Yukawa triplet diagonal, $U_L^\dagger\mathcal Y_aU_R=D_a$, then
\begin{equation}
\mathcal Y_a\mathcal Y_b^\dagger=U_LD_aD_b^\dagger U_L^\dagger.
\label{eq:flavour-product-condition}
\end{equation}
Every such product must therefore be normal, and all products must commute. These necessary conditions suffice to exclude the tensor family in this paper. An invertible change of scalar coordinates changes only the basis of its linear span, so we can use the coefficients of $(c,p,q)$ instead of the original three family components.

For $\mu\ne0$, divide by the common factor $\mu$ and put $r=\lambda/\mu$, $g=r+1$ and $b=2r+1$. The coefficient matrices are
\begin{equation}
\begin{gathered}
A=\operatorname{diag}(2g,b,b),\qquad
B=\begin{pmatrix}0&g&0\\g&0&1/2\\0&1/2&0\end{pmatrix},\\
C=\begin{pmatrix}0&0&g\\0&1/2&0\\g&0&-1/2\end{pmatrix}.
\end{gathered}
\label{eq:flavour-abc}
\end{equation}
Write $r=x+iy$, where $x,y$ are real, and set $X=BA^\dagger$. Two entries of its normality defect are
\begin{equation}
\begin{aligned}
\bigl[X,X^\dagger\bigr]_{11}&=-(4x+3)|g|^2,\\
\bigl[X,X^\dagger\bigr]_{13}&=\frac12g\bar b(-1+4iy).
\end{aligned}
\label{eq:flavour-normality-defect}
\end{equation}
Since $|-1+4iy|^2=1+16y^2>0$, normality requires $g=0$ or $b=0$, without division by either factor. On the branch $b=0$, $r=-1/2$ and the first defect is $-1/4$, so normality fails. On the branch $g=0$, $r=-1$; with $Y=CA^\dagger$ one obtains
\begin{equation}
[X,Y]_{23}=-\frac12,
\label{eq:flavour-exceptional-branch}
\end{equation}
so the product family fails to commute. This covers every finite complex $r$, including all degenerate mass branches. At the remaining projective point $\mu=0$, scale $\lambda$ to one and let $E_{ij}$ denote the $3\times3$ matrix units. Then $A=2I$, $B=E_{12}+E_{21}$ and $C=E_{13}+E_{31}$, giving
\begin{equation}
[BA^\dagger,CA^\dagger]=4(E_{23}-E_{32})\ne0.
\label{eq:flavour-pure-original}
\end{equation}
Thus no nonzero coupling pair admits a common biunitary diagonalisation; the zero pair gives the zero triplet. Conversely, any common diagonal basis after a scalar-coordinate change would diagonalise its inverse linear combinations, proving scalar-basis independence. The argument applies after the specified fermion canonical normalisation and separately to each charge sector.

For the explicit rotation in Eq.~\eqref{eq:scalar-higgs-rotation}, the signs of the real fermion eigenvectors are fixed by making their largest-magnitude components positive, and their columns are ordered by increasing absolute eigenvalue. Together with the right-handed signs in Eq.~\eqref{eq:flavour-mass-basis}, this gives the following additional neutral couplings at $M_Z$:
\begin{align}
\Gamma_2^e&=10^{-3}\begin{pmatrix}4.97458&4.53069&1.49812\\4.53069&4.77609&1.87690\\1.49812&1.87690&0.780604\end{pmatrix},\nonumber\\
\Gamma_3^e&=10^{-3}\begin{pmatrix}6.70403&-4.15734&1.81360\\-4.15734&3.06992&-1.53602\\1.81360&-1.53602&0.894314\end{pmatrix},\nonumber\\
\Gamma_2^d&=10^{-3}\begin{pmatrix}-4.89175&7.04014&2.08868\\-7.04014&10.6737&3.37893\\-2.08868&3.37893&1.02932\end{pmatrix},\nonumber\\
\Gamma_3^d&=10^{-3}\begin{pmatrix}-13.2700&-5.30091&3.59005\\5.30091&2.32051&-1.61070\\-3.59005&-1.61070&1.22015\end{pmatrix}.
\label{eq:flavour-gamma-values}
\end{align}
The down matrices are not symmetric because their first columns include the right-handed sign. These rounded matrices illustrate the couplings; the numerical amplitudes use the refined root of the certified mass system.

The matching can be derived directly from real-scalar currents. For a real neutral field $\varphi_\alpha$ with mass $m_\alpha$ and current $J_\alpha=\bar f_LK_\alpha f_R+\mathrm{h.c.}$, its mass and interaction terms are $-m_\alpha^2\varphi_\alpha^2/2-\varphi_\alpha J_\alpha$. Solving its low-momentum equation of motion and substituting back yields
\begin{equation}
\mathcal L_{\mathrm{eff}}=\sum_\alpha\frac{J_\alpha^2}{2m_\alpha^2},\qquad
\mathcal H_{\mathrm{eff}}=-\mathcal L_{\mathrm{eff}}.
\label{eq:flavour-current-matching}
\end{equation}
For $Q_{RR}^{ij}=(\bar f_iP_Rf_j)^2$, $Q_{LL}^{ij}=(\bar f_iP_Lf_j)^2$ and $Q_4^{ij}$, the Lagrangian coefficients are
\begin{equation}
\begin{gathered}
C_{RR}^{ij}=\sum_\alpha\frac{(K_\alpha)_{ij}^2}{2m_\alpha^2},\qquad
C_{LL}^{ij}=\sum_\alpha\frac{[(K_\alpha)_{ji}^*]^2}{2m_\alpha^2},\\
C_4^{ij}=\sum_\alpha\frac{(K_\alpha)_{ij}(K_\alpha)_{ji}^*}{m_\alpha^2}.
\end{gathered}
\label{eq:flavour-general-matching}
\end{equation}
In the real CP basis, $K_{H_\alpha}=\Gamma_\alpha/\sqrt2$ and $K_{A_\alpha}=i\Gamma_\alpha/\sqrt2$. Hence
\begin{align}
C_{RR}^{ij}&=\frac14\sum_{\alpha=2,3}(\Gamma_\alpha)_{ij}^2\left(\frac1{m_{H_\alpha}^2}-\frac1{m_{A_\alpha}^2}\right),\nonumber\\
C_{LL}^{ij}&=\frac14\sum_{\alpha=2,3}[(\Gamma_\alpha)_{ji}^*]^2\left(\frac1{m_{H_\alpha}^2}-\frac1{m_{A_\alpha}^2}\right),\nonumber\\
C_4^{ij}&=\frac12\sum_{\alpha=2,3}(\Gamma_\alpha)_{ij}(\Gamma_\alpha)_{ji}^*\left(\frac1{m_{H_\alpha}^2}+\frac1{m_{A_\alpha}^2}\right).
\label{eq:flavour-cp-matching}
\end{align}
The relative signs follow from $i^2=-1$ in the equal-chirality terms and $i(-i)=1$ in the opposite-chirality term. For $m_{H_\alpha}=m_{A_\alpha}=M$, Eq.~\eqref{eq:flavour-cp-matching} reduces to Eq.~\eqref{eq:flavour-lr-matching}. The aligned light Higgs has no flavour off-diagonal current. Summing the neutral spectrum is essential in multi-doublet flavour analyses~\cite{BabuWuXu2024}.

There is also a useful basis-independent expression. Put $\widehat{\mathcal Y}_a=U_L^\dagger\mathcal Y_aU_R$ and $\mathsf P_\perp=I-nn^T$. Completeness of the two directions orthogonal to the vacuum gives
\begin{equation}
\kappa_{ij}=\sum_{a,b}(\mathsf P_\perp)_{ab}(\widehat{\mathcal Y}_a)_{ij}(\widehat{\mathcal Y}_b)_{ji}^*.
\label{eq:flavour-completeness}
\end{equation}
Under a complex rotation of the degenerate doublets with zero VEVs, $\mathcal H'_\alpha=\sum_\beta U_{\alpha\beta}\mathcal H_\beta$, the Yukawa matrices transform as $\Gamma'_\alpha=\sum_\beta U_{\alpha\beta}^*\Gamma_\beta$. Unitarity preserves the summed contraction. A complex scalar rotation can mix the fields called CP even and CP odd in the original real basis, so this invariance belongs to the complete exchange. Fermion rephasings transform the operators along with their coefficients; a single complex Wilson coefficient need not be invariant by itself.

We next specify the QCD and hadronic conventions used in the finite example. The electroweak interface is $\mu_E=160\,\mathrm{GeV}$, with five active quarks between $M_Z=91.1876\,\mathrm{GeV}$ and $\mu_E$ and six above $\mu_E$. We take $\alpha_s^{(5)}(M_Z)=0.1181$. Leading QCD gives
\begin{equation}
\begin{gathered}
\frac{d\alpha_s}{d\log\mu}=-\frac{\beta_0\alpha_s^2}{2\pi},\qquad \beta_0=11-\frac{2n_f}{3},\\
\frac{d\mathcal Y_{da}}{d\log\mu}=-8\frac{\alpha_s}{4\pi}\mathcal Y_{da}.
\end{gathered}
\label{eq:flavour-qcd-equations}
\end{equation}
which rescales all down-sector entries identically and leaves the fermion singular vectors unchanged. Writing $\alpha_Z$, $\alpha_E$ and $\alpha_M$ for the strong coupling at the three scales,
\begin{equation}
\begin{gathered}
\mathcal Y_{da}(M)=r_m\mathcal Y_{da}(M_Z),\\
r_m=\left(\frac{\alpha_E}{\alpha_Z}\right)^{4/\beta_0^{(5)}}
\left(\frac{\alpha_M}{\alpha_E}\right)^{4/\beta_0^{(6)}}.
\end{gathered}
\label{eq:flavour-yukawa-running}
\end{equation}

In the BMU basis~\cite{BurasJagerUrban2001},
\begin{equation}
\begin{gathered}
Q_1^{LR}=(\bar d_i\gamma_\mu P_Ld_j)(\bar d_i\gamma^\mu P_Rd_j),\qquad Q_2^{LR}=Q_4^{ij},\\
\gamma_{LR}^{(0)}=\begin{pmatrix}2&12\\0&-16\end{pmatrix},\qquad
\frac{dW}{d\log\mu}=\frac{\alpha_s}{4\pi}(\gamma_{LR}^{(0)})^TW.
\end{gathered}
\label{eq:flavour-bmu-running}
\end{equation}
Tree matching gives $W_1(M)=0$. At this order it remains zero, while
\begin{equation}
U_{22}(\mu_E,M)=\left(\frac{\alpha_M}{\alpha_E}\right)^{-8/\beta_0^{(6)}}.
\label{eq:flavour-u22}
\end{equation}
The two Yukawa factors at matching therefore combine with Wilson running to give
\begin{equation}
r_m^2U_{22}=\left(\frac{\alpha_E}{\alpha_Z}\right)^{8/\beta_0^{(5)}}=0.9219213271.
\label{eq:flavour-qcd-net}
\end{equation}
This exact cancellation of the above-interface leading logarithms holds within the stated QCD approximation. At $M=2\times10^6\,\mathrm{GeV}$, the separate factors are $r_m=0.62028852$ and $U_{22}=2.39610793$.

The lower-scale input is the LEFT master formula of Ref.~\cite{AebischerEtAl2020Atlas}, written at the common NDR interface as
\begin{equation}
2M_{12}^{P,\mathrm{new}}=\Delta M_P^{\mathrm{exp}}\sum_{a=1,2}P_a^{LR}(\mu_E)W_a^{LR}(\mu_E).
\label{eq:flavour-hadronic-interface}
\end{equation}
The $P$ factors include low-scale evolution, matrix elements and chiral enhancement. In the order $(K,B_d,B_s)$, their values in $\mathrm{TeV}^2$ are
\begin{equation}
\begin{aligned}
(P_1^{LR})&=(-8.55\times10^7,-1.029\times10^6,-4.20\times10^4),\\
(P_2^{LR})&=(1.414\times10^8,1.275\times10^6,5.22\times10^4).
\end{aligned}
\label{eq:flavour-p-values}
\end{equation}
The quoted errors on $P_2^{LR}$ are $(8.2\times10^6,5.0\times10^4,2.1\times10^3)\,\mathrm{TeV}^2$. The accompanying reference values are $\Delta M_P^{\mathrm{exp}}=(3.484\times10^{-15},3.334\times10^{-13},1.1683\times10^{-11})\,\mathrm{GeV}$. These fixed inputs define the example. No further low-scale quark masses should be inserted into Eq.~\eqref{eq:flavour-hadronic-interface}. For Wilson coefficients in $\mathrm{GeV}^{-2}$,
\begin{equation}
\mathcal R_P=10^6\left|\sum_a P_a^{LR}[\mathrm{TeV}^2]W_a^{LR}[\mathrm{GeV}^{-2}]\right|,
\label{eq:flavour-unit-conversion}
\end{equation}
and the leading scalar result implies
\begin{equation}
M_{\mathrm{lower}}[\mathrm{TeV}]=\sqrt{\frac{|\kappa_{ij}(M_Z)|\,(0.9219213271)\,P_2^{LR}[\mathrm{TeV}^2]}{0.1}}.
\label{eq:flavour-leading-lower}
\end{equation}
This gives $(318.18,14.24,2.597)\,\mathrm{TeV}$ for $(K,B_d,B_s)$.

For the running diagnostic we include the two-loop beta coefficient $\beta_1=102-38n_f/3$, the mass anomalous dimension $\gamma_m^{(1)}=404/3-40n_f/9$~\cite{BaikovChetyrkinKuhn2014}, and the NDR matrix~\cite{BurasJagerUrban2001}
\begin{equation}
\gamma_{LR}^{(1)}=\begin{pmatrix}71/3-22n_f/9&198-44n_f/3\\225/4-2n_f&-1343/6+68n_f/9\end{pmatrix}.
\label{eq:flavour-two-loop-matrix}
\end{equation}
With $a_s=\alpha_s/(4\pi)$, the conventions are $da_s/d\log\mu=-2\beta_0a_s^2-2\beta_1a_s^3$, $d\mathcal Y_{da}/d\log\mu=-(8a_s+\gamma_m^{(1)}a_s^2)\mathcal Y_{da}$, and $dW/d\log\mu=[a_s\gamma_{LR}^{(0)}+a_s^2\gamma_{LR}^{(1)}]^TW$. Retaining tree matching gives, at the finite point,
\begin{equation}
r_m^2U_{12}=-0.00861304,\qquad r_m^2U_{22}=0.90443610.
\label{eq:flavour-two-loop-values}
\end{equation}
We use $\mathcal R_{P,\mathrm{diag,abs}}=10^6\sum_a|P_a^{LR}W_{a,\mathrm{diag}}^{LR}|$ to avoid cancellations between its two operator contributions. The conditional amplitude test uses $8\max(\mathcal R_{P,\mathrm{LO}},\mathcal R_{P,\mathrm{diag,abs}})$. Recomputing the running when solving for its lower boundary gives $(899.96,40.28,7.345)\,\mathrm{TeV}$. At $2000\,\mathrm{TeV}$ the corresponding fractions are $(0.0202483,4.05554\times10^{-5},1.34859\times10^{-6})$. The factor eight combines an assumed factor two in the hadronic amplitudes and a factor four in the short-distance amplitudes; these are assigned allowances, rather than calculated bounds on missing processes. Finite NLO matching, electroweak and Yukawa running are outside this diagnostic; evolving this pure-QCD subsector above electroweak breaking does not supply full SMEFT evolution.

Finally, the surviving contact interaction relevant to $\mu\to3e$ is
\begin{equation}
\mathcal L_{\mu3e}=a(\bar eP_R\mu)(\bar eP_Le)+b(\bar eP_L\mu)(\bar eP_Re)+\mathrm{h.c.},
\label{eq:flavour-lepton-contact}
\end{equation}
with $a,b$ defined in Eq.~\eqref{eq:flavour-mu3e}. Fierz rearrangement relates these terms to the mixed-chirality vector operators of Ref.~\cite{KunoOkada2001}, with $|g_{5,6}|=|a,b|/(4\sqrt2G_F)$, up to relabelling and the conjugation appropriate to $\mu^+$ decay. Their contact branching ratio is $|g_5|^2+|g_6|^2$.

The same normalisation follows by direct integration. Let $p_1,p_2$ be the two electron momenta and $p_3$ the positron momentum in $\mu^-$ decay. At $m_e=0$, spin averaging and antisymmetrisation give
\begin{equation}
\overline{|\mathcal A_a|^2}=2|a|^2\bigl[(p_\mu\!\cdot p_1)(p_2\!\cdot p_3)+(p_\mu\!\cdot p_2)(p_1\!\cdot p_3)\bigr].
\label{eq:flavour-lepton-square}
\end{equation}
The interference of the exchanged diagrams and the $a$--$b$ interference vanish in this limit. For $s=(p_2+p_3)^2/m_\mu^2$ and $t=(p_1+p_3)^2/m_\mu^2$, the allowed triangle is $s,t\ge0$, $s+t\le1$, and
\begin{equation}
\int_0^1 ds\int_0^{1-s}dt\,\frac{s(1-s)+t(1-t)}2=\frac1{12}.
\label{eq:flavour-lepton-phase-space}
\end{equation}
Including the identical-electron factor $1/2!$ yields
\begin{equation}
\Gamma(\mu\to3e)=\frac{m_\mu^5(|a|^2+|b|^2)}{6144\pi^3}.
\label{eq:flavour-lepton-rate}
\end{equation}
Dividing by the leading muon decay width $G_F^2m_\mu^5/(192\pi^3)$ gives Eq.~\eqref{eq:flavour-mu3e}. With the benchmark couplings, $M^2a=M^2b=-5.33265123\times10^{-6}$, while replacing each orthogonal-direction sum by the sum of absolute values gives $5.04091395\times10^{-5}$. A further factor four in each amplitude gives a branching ratio of $1.17\times10^{-24}$. This contact contribution is compared with the SINDRUM limit~\cite{BellgardtEtAl1988}; photon dipoles, QED evolution and other lepton-flavour processes are not included.

\section*{Declaration of generative AI and AI-assisted technologies in the manuscript preparation process}

During the preparation of this manuscript, OpenAI ChatGPT (GPT-6 models) was used to assist with mathematical exploration, code development, literature searches and manuscript editing. The authors reviewed and verified the AI-assisted content and take full responsibility for the results and conclusions.

\bibliography{referencev2}

@article{GillardGresnigt2019,
  author        = {Gillard, Adam B. and Gresnigt, Niels G.},
  title         = {Three fermion generations with two unbroken gauge symmetries from the complex sedenions},
  journal       = {Eur. Phys. J. C},
  volume        = {79},
  number        = {5},
  pages         = {446},
  year          = {2019},
  doi           = {10.1140/epjc/s10052-019-6967-1},
  eprint        = {1904.03186},
  archivePrefix = {arXiv},
  primaryClass  = {hep-th}
}

@article{GresnigtGourlayVarma2023,
  author        = {Gresnigt, Niels and Gourlay, Liam and Varma, Abhinav},
  title         = {Three generations of colored fermions with {$S_3$} family symmetry from {Cayley--Dickson} sedenions},
  journal       = {Eur. Phys. J. C},
  volume        = {83},
  number        = {8},
  pages         = {747},
  year          = {2023},
  doi           = {10.1140/epjc/s10052-023-11923-y},
  eprint        = {2306.13098},
  archivePrefix = {arXiv},
  primaryClass  = {physics.gen-ph}
}

@article{GourlayGresnigt2024,
  author        = {Gourlay, Liam and Gresnigt, Niels},
  title         = {Algebraic realisation of three fermion generations with {$S_3$} family and unbroken gauge symmetry from {$\mathbb{C}\ell(8)$}},
  journal       = {Eur. Phys. J. C},
  volume        = {84},
  number        = {10},
  pages         = {1129},
  year          = {2024},
  doi           = {10.1140/epjc/s10052-024-13476-0},
  eprint        = {2407.01580},
  archivePrefix = {arXiv},
  primaryClass  = {hep-th}
}

@article{Gresnigt2026Generations,
  author        = {Gresnigt, Niels},
  title         = {Three fermion generations and an untriplicated gauge sector with {$S_3$} family symmetry},
  journal       = {Phys. Lett. B},
  volume        = {880},
  pages         = {140782},
  year          = {2026},
  doi           = {10.1016/j.physletb.2026.140782}
}

@misc{Gresnigt2026Technical,
  author        = {Gresnigt, Niels},
  title         = {Electroweak Structure and Three Fermion Generations in {Clifford} Algebra with {$S_3$} Family Symmetry},
  year          = {2026},
  eprint        = {2601.07857},
  archivePrefix = {arXiv},
  primaryClass  = {physics.gen-ph},
  note          = {Version 2},
  url           = {https://arxiv.org/abs/2601.07857v2}
}

@misc{Gresnigt2026Higgs,
  author        = {Gresnigt, Niels},
  title         = {{Higgs} and {Yukawa} Structure in a {Clifford} Algebra Model with Three Generations and {$S_3$} Family Symmetry},
  year          = {2026},
  eprint        = {2604.24795},
  archivePrefix = {arXiv},
  primaryClass  = {physics.gen-ph},
  note          = {Version 2},
  url           = {https://arxiv.org/abs/2604.24795v2}
}

@article{Gresnigt2026Review,
  author        = {Gresnigt, Niels},
  title         = {Sedenions, {Clifford} Algebras, and Three Fermion Generations: A Focused Review},
  journal       = {Fortschr. Phys.},
  volume        = {74},
  number        = {11},
  pages         = {e70161},
  year          = {2026},
  doi           = {10.1002/prop.70161},
  eprint        = {2609.10569},
  archivePrefix = {arXiv},
  primaryClass  = {physics.gen-ph}
}

@article{Brown1967,
  author        = {Brown, Robert},
  title         = {On generalized {Cayley--Dickson} algebras},
  journal       = {Pacific J. Math.},
  volume        = {20},
  number        = {3},
  pages         = {415--422},
  year          = {1967},
  doi           = {10.2140/pjm.1967.20.415}
}

@article{ManogueDray1999,
  author        = {Manogue, Corinne A. and Dray, Tevian},
  title         = {Dimensional reduction},
  journal       = {Mod. Phys. Lett. A},
  volume        = {14},
  number        = {2},
  pages         = {99--104},
  year          = {1999},
  doi           = {10.1142/S0217732399000134},
  eprint        = {hep-th/9807044},
  archivePrefix = {arXiv},
  primaryClass  = {hep-th}
}

@article{Dixon2004FamilyReplication,
  author        = {Dixon, Geoffrey},
  title         = {Division algebras: Family replication},
  journal       = {J. Math. Phys.},
  volume        = {45},
  number        = {10},
  pages         = {3878--3882},
  year          = {2004},
  doi           = {10.1063/1.1786682}
}

@article{GordingSchmidtMay2020,
  author        = {Gording, Brage and Schmidt-May, Angnis},
  title         = {The Unified {Standard Model}},
  journal       = {Adv. Appl. Clifford Algebras},
  volume        = {30},
  number        = {4},
  pages         = {55},
  year          = {2020},
  doi           = {10.1007/s00006-020-01082-8},
  eprint        = {1909.05641},
  archivePrefix = {arXiv},
  primaryClass  = {hep-th}
}

@article{Furey2018Generations,
  author        = {Furey, C.},
  title         = {Three generations, two unbroken gauge symmetries, and one eight-dimensional algebra},
  journal       = {Phys. Lett. B},
  volume        = {785},
  pages         = {84--89},
  year          = {2018},
  doi           = {10.1016/j.physletb.2018.08.032},
  eprint        = {1910.08395},
  archivePrefix = {arXiv},
  primaryClass  = {hep-th}
}

@article{IshimoriEtAl2010,
  author        = {Ishimori, Hajime and Kobayashi, Tatsuo and Ohki, Hiroshi and Shimizu, Yusuke and Okada, Hiroshi and Tanimoto, Morimitsu},
  title         = {{Non-Abelian} discrete symmetries in particle physics},
  journal       = {Prog. Theor. Phys. Suppl.},
  volume        = {183},
  pages         = {1--163},
  year          = {2010},
  doi           = {10.1143/PTPS.183.1},
  eprint        = {1003.3552},
  archivePrefix = {arXiv},
  primaryClass  = {hep-th}
}

@article{BrancoEtAl2012,
  author        = {Branco, G. C. and Ferreira, P. M. and Lavoura, L. and Rebelo, M. N. and Sher, Marc and Silva, Jo{\~a}o P.},
  title         = {Theory and phenomenology of two-{Higgs}-doublet models},
  journal       = {Phys. Rept.},
  volume        = {516},
  pages         = {1--102},
  year          = {2012},
  doi           = {10.1016/j.physrep.2012.02.002},
  eprint        = {1106.0034},
  archivePrefix = {arXiv},
  primaryClass  = {hep-ph}
}

@article{GlashowWeinberg1977,
  author        = {Glashow, Sheldon L. and Weinberg, Steven},
  title         = {Natural conservation laws for neutral currents},
  journal       = {Phys. Rev. D},
  volume        = {15},
  pages         = {1958--1965},
  year          = {1977},
  doi           = {10.1103/PhysRevD.15.1958}
}

@article{Paschos1977,
  author        = {Paschos, Emmanuel A.},
  title         = {Diagonal neutral currents},
  journal       = {Phys. Rev. D},
  volume        = {15},
  pages         = {1966--1972},
  year          = {1977},
  doi           = {10.1103/PhysRevD.15.1966}
}

@article{PichTuzon2009,
  author        = {Pich, Antonio and Tuz{\'o}n, Paula},
  title         = {{Yukawa} alignment in the two-{Higgs}-doublet model},
  journal       = {Phys. Rev. D},
  volume        = {80},
  pages         = {091702},
  year          = {2009},
  doi           = {10.1103/PhysRevD.80.091702},
  eprint        = {0908.1554},
  archivePrefix = {arXiv},
  primaryClass  = {hep-ph}
}

@article{FerreiraLavouraSilva2010,
  author        = {Ferreira, P. M. and Lavoura, L. and Silva, Jo{\~a}o P.},
  title         = {Renormalization-group constraints on {Yukawa} alignment in multi-{Higgs}-doublet models},
  journal       = {Phys. Lett. B},
  volume        = {688},
  pages         = {341--344},
  year          = {2010},
  doi           = {10.1016/j.physletb.2010.04.033},
  eprint        = {1001.2561},
  archivePrefix = {arXiv},
  primaryClass  = {hep-ph}
}

@article{PenuelasPich2017,
  author        = {Pe{\~n}uelas, Ana and Pich, Antonio},
  title         = {Flavour alignment in multi-{Higgs}-doublet models},
  journal       = {J. High Energy Phys.},
  volume        = {2017},
  number        = {12},
  pages         = {084},
  year          = {2017},
  doi           = {10.1007/JHEP12(2017)084},
  eprint        = {1710.02040},
  archivePrefix = {arXiv},
  primaryClass  = {hep-ph}
}

@article{RodejohannSaldanaSalazar2019,
  author        = {Rodejohann, Werner and Salda{\~n}a-Salazar, Ulises},
  title         = {Multi-{Higgs}-doublet models and singular alignment},
  journal       = {J. High Energy Phys.},
  volume        = {2019},
  number        = {07},
  pages         = {036},
  year          = {2019},
  doi           = {10.1007/JHEP07(2019)036},
  eprint        = {1903.00983},
  archivePrefix = {arXiv},
  primaryClass  = {hep-ph}
}

@article{BaekEtAl2024,
  author        = {Baek, S. and Kersten, J{\"o}rn and Ko, P. and Velasco-Sevilla, Liliana},
  title         = {An unfamiliar way to generate the hierarchy of {Standard Model} fermion masses},
  journal       = {J. High Energy Phys.},
  volume        = {2024},
  number        = {02},
  pages         = {143},
  year          = {2024},
  doi           = {10.1007/JHEP02(2024)143},
  eprint        = {2309.07788},
  archivePrefix = {arXiv},
  primaryClass  = {hep-ph}
}

@article{VarzielasTalbert2020,
  author        = {de Medeiros Varzielas, Ivo and Talbert, Jim},
  title         = {{FCNC}-free multi-{Higgs}-doublet models from broken family symmetries},
  journal       = {Phys. Lett. B},
  volume        = {800},
  pages         = {135091},
  year          = {2020},
  doi           = {10.1016/j.physletb.2019.135091},
  eprint        = {1908.10979},
  archivePrefix = {arXiv},
  primaryClass  = {hep-ph}
}

@article{PakvasaSugawara1978,
  author        = {Pakvasa, Sandip and Sugawara, Hirotaka},
  title         = {Discrete symmetry and {Cabibbo} angle},
  journal       = {Phys. Lett. B},
  volume        = {73},
  pages         = {61--64},
  year          = {1978},
  doi           = {10.1016/0370-2693(78)90172-7}
}

@article{KuboEtAl2003,
  author        = {Kubo, Jisuke and Mondrag{\'o}n, Alfonso and Mondrag{\'o}n, Myriam and Rodr{\'i}guez-J{\'a}uregui, Ezequiel},
  title         = {The flavor symmetry},
  journal       = {Prog. Theor. Phys.},
  volume        = {109},
  pages         = {795--807},
  year          = {2003},
  doi           = {10.1143/PTP.109.795},
  eprint        = {hep-ph/0302196},
  archivePrefix = {arXiv},
  primaryClass  = {hep-ph},
  note          = {Erratum: Prog. Theor. Phys. 114, 287 (2005)}
}

@article{KuboOkadaSakamaki2004,
  author        = {Kubo, Jisuke and Okada, Hiroshi and Sakamaki, Fumiaki},
  title         = {{Higgs} potential in a minimal {$S_3$} invariant extension of the {Standard Model}},
  journal       = {Phys. Rev. D},
  volume        = {70},
  pages         = {036007},
  year          = {2004},
  doi           = {10.1103/PhysRevD.70.036007},
  eprint        = {hep-ph/0402089},
  archivePrefix = {arXiv},
  primaryClass  = {hep-ph}
}

@article{GonzalezCanalesEtAl2013,
  author        = {Gonz{\'a}lez Canales, F. and Mondrag{\'o}n, A. and Mondrag{\'o}n, M. and Salda{\~n}a Salazar, U. J. and Velasco-Sevilla, L.},
  title         = {Quark sector of {$S_3$} models: Classification and comparison with experimental data},
  journal       = {Phys. Rev. D},
  volume        = {88},
  pages         = {096004},
  year          = {2013},
  doi           = {10.1103/PhysRevD.88.096004},
  eprint        = {1304.6644},
  archivePrefix = {arXiv},
  primaryClass  = {hep-ph}
}

@article{DasDey2014,
  author        = {Das, Dipankar and Dey, Ujjal Kumar},
  title         = {Analysis of an extended scalar sector with {$S_3$} symmetry},
  journal       = {Phys. Rev. D},
  volume        = {89},
  pages         = {095025},
  year          = {2014},
  doi           = {10.1103/PhysRevD.89.095025},
  eprint        = {1404.2491},
  archivePrefix = {arXiv},
  primaryClass  = {hep-ph},
  note          = {Erratum: Phys. Rev. D 91, 039905 (2015)}
}

@article{EmmanuelCostaEtAl2016,
  author        = {Emmanuel-Costa, D. and Ogreid, O. M. and Osland, P. and Rebelo, M. N.},
  title         = {Spontaneous symmetry breaking in the {$S_3$}-symmetric scalar sector},
  journal       = {J. High Energy Phys.},
  volume        = {2016},
  number        = {02},
  pages         = {154},
  year          = {2016},
  doi           = {10.1007/JHEP02(2016)154},
  eprint        = {1601.04654},
  archivePrefix = {arXiv},
  primaryClass  = {hep-ph},
  note          = {Erratum: J. High Energy Phys. 2016 (08), 169}
}

@article{Chakrabarty2016,
  author        = {Chakrabarty, Nabarun},
  title         = {High-scale validity of a model with three {Higgs} doublets},
  journal       = {Phys. Rev. D},
  volume        = {93},
  pages         = {075025},
  year          = {2016},
  doi           = {10.1103/PhysRevD.93.075025},
  eprint        = {1511.08137},
  archivePrefix = {arXiv},
  primaryClass  = {hep-ph}
}

@article{GomezBockEtAl2021,
  author        = {G{\'o}mez-Bock, M. and Mondrag{\'o}n, M. and P{\'e}rez-Mart{\'i}nez, A.},
  title         = {Scalar and gauge sectors in the {3-Higgs Doublet Model} under the {$S_3$} symmetry},
  journal       = {Eur. Phys. J. C},
  volume        = {81},
  pages         = {942},
  year          = {2021},
  doi           = {10.1140/epjc/s10052-021-09731-3},
  eprint        = {2102.02800},
  archivePrefix = {arXiv},
  primaryClass  = {hep-ph}
}

@article{BabuWuXu2024,
  author        = {Babu, K. S. and Wu, Yongcheng and Xu, Shiyuan},
  title         = {Fermion masses, neutrino mixing and {Higgs}-mediated flavor violation in {3HDM} with {$S_3$} permutation symmetry},
  journal       = {J. High Energy Phys.},
  volume        = {2024},
  number        = {12},
  pages         = {166},
  year          = {2024},
  doi           = {10.1007/JHEP12(2024)166},
  eprint        = {2312.15828},
  archivePrefix = {arXiv},
  primaryClass  = {hep-ph}
}

@article{KingPeddie2004,
  author        = {King, S. F. and Peddie, I. N. R.},
  title         = {Canonical normalisation and {Yukawa} matrices},
  journal       = {Phys. Lett. B},
  volume        = {586},
  pages         = {83--94},
  year          = {2004},
  doi           = {10.1016/j.physletb.2004.01.087},
  eprint        = {hep-ph/0312237},
  archivePrefix = {arXiv},
  primaryClass  = {hep-ph}
}

@article{EspinosaIbarra2004,
  author        = {Espinosa, J. R. and Ibarra, A.},
  title         = {Flavour symmetries and {K\"ahler} operators},
  journal       = {J. High Energy Phys.},
  volume        = {2004},
  number        = {08},
  pages         = {010},
  year          = {2004},
  doi           = {10.1088/1126-6708/2004/08/010},
  eprint        = {hep-ph/0405095},
  archivePrefix = {arXiv},
  primaryClass  = {hep-ph}
}

@article{ChenEtAl2012,
  author        = {Chen, M.-C. and Fallbacher, M. and Ratz, M. and Staudt, C.},
  title         = {On predictions from spontaneously broken flavor symmetries},
  journal       = {Phys. Lett. B},
  volume        = {718},
  pages         = {516--521},
  year          = {2012},
  doi           = {10.1016/j.physletb.2012.10.077},
  eprint        = {1208.2947},
  archivePrefix = {arXiv},
  primaryClass  = {hep-ph}
}

@article{Kawamura2019,
  author        = {Kawamura, Y.},
  title         = {{Yukawa} interactions, flavor symmetry, and non-canonical {K\"ahler} potential},
  journal       = {Prog. Theor. Exp. Phys.},
  volume        = {2019},
  number        = {4},
  pages         = {043B05},
  year          = {2019},
  doi           = {10.1093/ptep/ptz018},
  eprint        = {1808.06315},
  archivePrefix = {arXiv},
  primaryClass  = {hep-ph}
}

@article{KakizakiYamaguchi2003,
  author        = {Kakizaki, M. and Yamaguchi, M.},
  title         = {Singular {K\"ahler} potential and heavy top quark in a democratic mass matrix model},
  journal       = {Phys. Lett. B},
  volume        = {573},
  pages         = {123--130},
  year          = {2003},
  doi           = {10.1016/j.physletb.2003.08.040},
  eprint        = {hep-ph/0307362},
  archivePrefix = {arXiv},
  primaryClass  = {hep-ph}
}

@article{XingZhangZhou2012,
  author        = {Xing, Z.-z. and Zhang, H. and Zhou, S.},
  title         = {Impacts of the {Higgs} mass on vacuum stability, running fermion masses, and two-body {Higgs} decays},
  journal       = {Phys. Rev. D},
  volume        = {86},
  pages         = {013013},
  year          = {2012},
  doi           = {10.1103/PhysRevD.86.013013},
  eprint        = {1112.3112},
  archivePrefix = {arXiv},
  primaryClass  = {hep-ph}
}

@article{AntuschHinzeSaad2026,
  author        = {Antusch, S. and Hinze, K. and Saad, S.},
  title         = {Updated running quark and lepton parameters at various scales},
  journal       = {Phys. Rev. D},
  volume        = {113},
  pages         = {095011},
  year          = {2026},
  doi           = {10.1103/fdcc-ycph},
  eprint        = {2510.01312},
  archivePrefix = {arXiv},
  primaryClass  = {hep-ph}
}

@article{PDG2026,
  author        = {Takahashi, F. and others},
  collaboration = {Particle Data Group},
  title         = {Review of Particle Physics},
  journal       = {Int. J. Mod. Phys. A},
  volume        = {41},
  number        = {22},
  pages         = {2630011},
  year          = {2026},
  doi           = {10.1142/S0217751X26300115}
}

@article{Youla1961,
  author  = {Youla, D. C.},
  title   = {A normal form for a matrix under the unitary congruence group},
  journal = {Canad. J. Math.},
  volume  = {13},
  pages   = {694--704},
  year    = {1961},
  doi     = {10.4153/CJM-1961-059-8}
}

@book{HornJohnson2012,
  author    = {Horn, Roger A. and Johnson, Charles R.},
  title     = {Matrix Analysis},
  edition   = {2},
  publisher = {Cambridge University Press},
  address   = {Cambridge},
  year      = {2013},
  isbn      = {9781139020411},
  doi       = {10.1017/CBO9781139020411}
}

@book{MooreKearfottCloud2009,
  author    = {Moore, Ramon E. and Kearfott, R. Baker and Cloud, Michael J.},
  title     = {Introduction to Interval Analysis},
  publisher = {Society for Industrial and Applied Mathematics},
  address   = {Philadelphia},
  year      = {2009},
  isbn      = {9780898716696},
  doi       = {10.1137/1.9780898717716}
}

@article{Rump2010,
  author  = {Rump, Siegfried M.},
  title   = {Verification methods: Rigorous results using floating-point arithmetic},
  journal = {Acta Numer.},
  volume  = {19},
  pages   = {287--449},
  year    = {2010},
  doi     = {10.1017/S096249291000005X}
}

@article{Neumaier1989,
  author  = {Neumaier, Arnold},
  title   = {Rigorous sensitivity analysis for parameter-dependent systems of equations},
  journal = {J. Math. Anal. Appl.},
  volume  = {144},
  number  = {1},
  pages   = {16--25},
  year    = {1989},
  doi     = {10.1016/0022-247X(89)90357-0}
}

@article{BurasJagerUrban2001,
  author = {Buras, Andrzej J. and J{\"a}ger, Sebastian and Urban, J{\"o}rg},
  title = {Master formulae for {$\Delta F=2$} {NLO-QCD} factors in the {Standard Model} and beyond},
  journal = {Nucl. Phys. B},
  volume = {605},
  pages = {600--624},
  year = {2001},
  doi = {10.1016/S0550-3213(01)00207-3},
  eprint = {hep-ph/0102316},
  archivePrefix = {arXiv},
  primaryClass = {hep-ph}
}

@article{AebischerEtAl2020Atlas,
  author = {Aebischer, Jason and Bobeth, Christoph and Buras, Andrzej J. and Kumar, Jacky},
  title = {{SMEFT} atlas of {$\Delta F=2$} transitions},
  journal = {J. High Energy Phys.},
  volume = {2020},
  number = {12},
  pages = {187},
  year = {2020},
  doi = {10.1007/JHEP12(2020)187},
  eprint = {2009.07276},
  archivePrefix = {arXiv},
  primaryClass = {hep-ph}
}

@article{BaikovChetyrkinKuhn2014,
  author = {Baikov, P. A. and Chetyrkin, K. G. and K{\"u}hn, J. H.},
  title = {Quark mass and field anomalous dimensions to {$\mathcal{O}(\alpha_s^5)$}},
  journal = {J. High Energy Phys.},
  volume = {2014},
  number = {10},
  pages = {076},
  year = {2014},
  doi = {10.1007/JHEP10(2014)076},
  eprint = {1402.6611},
  archivePrefix = {arXiv},
  primaryClass = {hep-ph}
}

@article{KunoOkada2001,
  author = {Kuno, Yoshitaka and Okada, Yasuhiro},
  title = {Muon decay and physics beyond the {Standard Model}},
  journal = {Rev. Mod. Phys.},
  volume = {73},
  pages = {151--202},
  year = {2001},
  doi = {10.1103/RevModPhys.73.151},
  eprint = {hep-ph/9909265},
  archivePrefix = {arXiv},
  primaryClass = {hep-ph}
}

@article{BellgardtEtAl1988,
  author = {Bellgardt, U. and others},
  collaboration = {SINDRUM},
  title = {Search for the decay {$\mu^+\to e^+e^+e^-$}},
  journal = {Nucl. Phys. B},
  volume = {299},
  number = {1},
  pages = {1--6},
  year = {1988},
  doi = {10.1016/0550-3213(88)90462-2}
}

@article{DarvishiMasouminiaPilaftsis2021,
  author = {Darvishi, Neda and Masouminia, M. R. and Pilaftsis, Apostolos},
  title = {Maximally symmetric three-{Higgs}-doublet model},
  journal = {Phys. Rev. D},
  volume = {104},
  number = {11},
  pages = {115017},
  year = {2021},
  doi = {10.1103/PhysRevD.104.115017},
  eprint = {2106.03159},
  archivePrefix = {arXiv},
  primaryClass = {hep-ph}
}

@article{VarzielasIvanovLevy2021,
  author = {de Medeiros Varzielas, Ivo and Ivanov, Igor P. and Levy, Miguel},
  title = {Exploring multi-{Higgs} models with softly broken large discrete symmetry groups},
  journal = {Eur. Phys. J. C},
  volume = {81},
  number = {10},
  pages = {918},
  year = {2021},
  doi = {10.1140/epjc/s10052-021-09681-w},
  eprint = {2107.08227},
  archivePrefix = {arXiv},
  primaryClass = {hep-ph}
}

@article{CarroloEtAl2021,
  author = {Carrolo, Sergio and Rom{\~a}o, Jorge C. and Silva, Jo{\~a}o P. and Vaz{\~a}o, Francisco},
  title = {Symmetry and decoupling in multi-{Higgs} boson models},
  journal = {Phys. Rev. D},
  volume = {103},
  number = {7},
  pages = {075026},
  year = {2021},
  doi = {10.1103/PhysRevD.103.075026},
  eprint = {2102.11303},
  archivePrefix = {arXiv},
  primaryClass = {hep-ph}
}

@article{KuncinasEtAl2020,
  author = {Kun{\v c}inas, A. and Osland, P. and Ogreid, O. M. and Rebelo, M. N.},
  title = {{$S_3$}-inspired three-{Higgs}-doublet models: A class with a complex vacuum},
  journal = {Phys. Rev. D},
  volume = {101},
  number = {7},
  pages = {075052},
  year = {2020},
  doi = {10.1103/PhysRevD.101.075052},
  eprint = {2001.01994},
  archivePrefix = {arXiv},
  primaryClass = {hep-ph}
}

@article{AltmannshoferToner2025,
  author = {Altmannshofer, Wolfgang and Toner, Kevin},
  title = {Flavor constraints in a generational three-{Higgs}-doublet model},
  journal = {Phys. Rev. D},
  volume = {111},
  number = {7},
  pages = {075009},
  year = {2025},
  doi = {10.1103/PhysRevD.111.075009},
  eprint = {2502.04579},
  archivePrefix = {arXiv},
  primaryClass = {hep-ph}
}

@article{kaneko2007flavor,
  author  = {Kaneko, Satoru and Sawanaka, Hideyuki and Shingai, Takaya and Tanimoto, Morimitsu and Yoshioka, Koichi},
  title   = {{Flavor symmetry and vacuum aligned mass textures}},
  journal = {Progress of Theoretical Physics},
  volume  = {117},
  number  = {1},
  pages   = {161--181},
  year    = {2007},
  doi     = {10.1143/PTP.117.161}
}

@incollection{Ablamowicz1995,
  author    = {Ab{\l}amowicz, Rafa{\l}},
  title     = {{Construction of Spinors via Witt Decomposition and Primitive Idempotents: A Review}},
  booktitle = {Clifford Algebras and Spinor Structures},
  editor    = {Ab{\l}amowicz, Rafa{\l} and Lounesto, Pertti},
  series    = {Mathematics and Its Applications},
  volume    = {321},
  pages     = {113--123},
  publisher = {Springer},
  address   = {Dordrecht},
  year      = {1995},
  doi       = {10.1007/978-94-015-8422-7_6}
}

@article{Gresnigt2020StandardModel,
  author        = {Gresnigt, Niels G.},
  title         = {{The {Standard Model} particle content with complete gauge symmetries from the minimal ideals of two {Clifford} algebras}},
  journal       = {The European Physical Journal C},
  volume        = {80},
  pages         = {583},
  year          = {2020},
  doi           = {10.1140/epjc/s10052-020-8141-1},
  eprint        = {2003.08814},
  archivePrefix = {arXiv},
  primaryClass  = {physics.gen-ph}
}

\end{document}